\documentclass[12pt,preprint]{aastex}
\usepackage{graphicx}
\usepackage[section] {placeins}
\usepackage{url}
\renewcommand\arraystretch{1.3}
\newcommand{\rmn}[1]{{\mathrm{#1}}}

\begin{document}

\title{The Exceptional Soft X-ray Halo of the Galaxy Merger NGC~6240}
\author{E.~Nardini\altaffilmark{1,2}, Junfeng Wang\altaffilmark{1,3}, 
G.~Fabbiano\altaffilmark{1}, M.~Elvis\altaffilmark{1}, S.~Pellegrini\altaffilmark{4}, 
G.~Risaliti\altaffilmark{1,5}, M.~Karovska\altaffilmark{1}, A.~Zezas\altaffilmark{1,6}}

\altaffiltext{1}{Harvard-Smithsonian Center for Astrophysics, 60 Garden Street, Cambridge, MA 02138, USA}
\altaffiltext{2}{\textit{Current Address:} Astrophysics Group, School of Physical and Geographical Sciences, 
Keele University, Keele, Staffordshire ST5 5BG, UK}
\altaffiltext{3}{\textit{Current Address:} Department of Physics and Astronomy and Center for Interdisciplinary 
Exploration and Research in Astronomy, Northwestern University, 2145 Sheridan Road, Evanston, IL 60208, USA}
\altaffiltext{4}{Dipartimento di Fisica e Astronomia, Universit\`a di Bologna, v.le Berti Pichat 6/2, 40127 Bologna, Italy}
\altaffiltext{5}{INAF - Osservatorio Astrofisico di Arcetri, L.go E. Fermi 5, 50125 Firenze, Italy}
\altaffiltext{6}{Physics Department, University of Crete, P.O. Box 2208, GR-710 03, Heraklion, Crete, Greece}
\email{e.nardini@keele.ac.uk}

\begin{abstract}
We report on a recent $\sim$150-ks long \textit{Chandra} observation of the ultraluminous 
infrared galaxy merger NGC~6240, which allows a detailed investigation of the diffuse galactic 
halo. Extended soft X-ray emission is detected at the 3$\sigma$ confidence level over a 
diamond-shaped region with projected physical size of $\sim$110$\times$80 kpc, and a  
single-component thermal model provides a reasonably good fit to the observed X-ray 
spectrum. The hot gas has a temperature of $\sim$7.5 million K, an estimated density of 
2.5$\times$10$^{-3}$ cm$^{-3}$, and a total mass of $\sim$10$^{10} M_{\sun}$, resulting 
in an intrinsic 0.4--2.5 keV luminosity of 4$\times$10$^{41}$ erg s$^{-1}$. The average 
temperature of 0.65 keV is quite high to be obviously related to either the binding energy in 
the dark-matter gravitational potential of the system or the energy dissipation and shocks 
following the galactic collision, yet the spatially-resolved spectral analysis reveals limited 
variations across the halo. The relative abundance of the main $\alpha$-elements with 
respect to iron is several times the solar value, and nearly constant as well, implying a 
uniform enrichment by type II supernovae out to the largest scales. Taken as a whole, the 
observational evidence is not compatible with a superwind originated by a recent, nuclear 
starburst, but rather hints at widespread, enhanced star formation proceeding at steady 
rate over the entire dynamical timescale ($\sim$200 Myr). The preferred scenario is that of a 
starburst-processed gas component gently expanding into, and mixing with, a pre-existing 
halo medium of lower metallicity ($Z \sim 0.1$ solar) and temperature ($kT \sim 0.25$ keV). 
This picture cannot be probed more extensively with the present data, and the ultimate fate 
of the diffuse, hot gas remains uncertain. Under some favorable conditions, at least a fraction 
of it might be retained after the merger completion, and evolve into the hot halo of a young 
elliptical galaxy. 
\end{abstract}


\section{Introduction}

NGC~6240 ($z \simeq 0.0245$; Downes et al. 1993) is one of the most impressive galaxy 
mergers in the local Universe. In the framework of hierarchical structure assembly, the 
remnant of the collision between two gas-rich spirals likely goes through a bright quasar 
phase, and eventually relaxes into a quiescent elliptical galaxy (e.g., Toomre \& Toomre 1972; 
Schweizer 1986; Barnes \& Hernquist 1992; Hopkins et al. 2008; and references therein). 
The gravitational instabilities at work during a major merger trigger an extensive gas 
redistribution, which consequently fuels both intense star formation and efficient black-hole 
accretion. Due to the large dust opacity to the intrinsic radiation field of this composite 
energy source (i.e., starburst and active galactic nucleus, AGN), the bulk of the bolometric 
luminosity possibly emerges at infrared wavelengths. The ultimate merging systems are thus 
found among the so-called Ultraluminous Infrared Galaxies (ULIRGs, $L_\rmn{IR} \sim 
L_\mathit{bol} > 10^{12} L_{\sun}$; Lonsdale et al. 2006). With an energy output of 
$\sim$7$\times$10$^{11} L_{\sun}$ at 8--1000~$\mu$m (e.g., Sanders \& Mirabel 1996), 
NGC~6240 falls slightly below this threshold, yet its global features are ordinary in ULIRGs, 
and it is usually included in this class (Genzel et al. 1998). 

The optical morphology of NGC~6240 is highly disturbed, with long tidal tails and a distorted 
disk obstructed by broad dust lanes (Fosbury \& Wall 1979; Gerssen et al. 2004). The central 
kpc harbors two nuclei, whose position and relative separation are wavelength-dependent 
owing to the significant obscuration (Schulz et al. 1993; Max et al. 2007). A huge amount 
(several $\times$10$^9 M_{\sun}$) of molecular gas is concentrated in between (Tacconi et 
al. 1999; Iono et al. 2007), giving rise to exceptional, shock-excited H$_2$ line emission in 
the near-IR (Joseph et al. 1984; Ohyama et al. 2003). The surrounding butterfly-shaped 
H$\alpha$ nebula shows filaments and loops, consistent with the expansion of superwind 
bubbles (Heckman et al. 1987; Lira et al. 2002). When not destroyed by the hard X-ray 
photons, also some dust is apparently entrained in these outflows (Bush et al. 2008). A soft 
X-ray component extended out to $\sim$10 kpc, and possibly beyond, was first detected by 
\textit{ROSAT} (Komossa et al. 1998), while \textit{Chandra} enabled the discovery of a buried 
AGN in each of the two nuclei (Komossa et al. 2003). This dual AGN character has been later 
supported by radio (Gallimore \& Beswick 2004) and mid-IR (Risaliti et al. 2006) observations. 

By virtue of its proximity and overall complexity, NGC~6240 is a unique source in various 
respects, whose study is extremely relevant to address many critical issues, including the 
transformation of galaxies through major mergers, the connection between AGN and 
starburst activity, the impact of AGN feedback and starburst superwinds, the chemical 
enrichment of the interstellar and intergalactic medium, the correlation between the 
host-galaxy bulges and their central supermassive black holes. In particular, as a dual AGN 
encounter close to the final coalescence, NGC~6240 represents a rare opportunity to directly 
witness a phenomenon thought to be common in the earlier cosmic epochs (Engel et al. 
2010a), but still not completely understood because of the dearth of observational constraints 
available in the present-day Universe. 

Taking advantage of the great improvement in data quality achieved through a recent 
\textit{Chandra} observation, here we explore the physical properties of the diffuse gas 
responsible for the large-scale ($r > 15\arcsec \simeq 7.5$ kpc), soft X-ray emission of 
NGC~6240, and investigate into its nature to extract some crucial information on the origin 
and evolution of the whole system. The luminosity of this $\sim$100-kpc wide, hot halo is 
quite high for normal early-type galaxies (Fabbiano et al. 1992), and more typical of the 
central objects in small groups and clusters (O'Sullivan et al. 2001a). If this were the case, 
the X-ray halo would be almost in virial equilibrium, and its luminosity would persist even 
after the merger completion. A companion paper (Wang et al. 2013a, in prep.) deals with the 
sub-arcsec resolution X-ray image of the nuclear region, focusing on the diffuse 5.5--8 
keV emission, and indicating the presence of fast shocks in the $\sim$70 million K hot gas 
phase. The detailed analysis of the soft X-ray emission in the intermediate nebular region 
($r < 15\arcsec$), which is pervaded by a starburst-driven wind, will be presented in a third, 
forthcoming paper (Wang et al. 2013b, in prep.). 

This work is organized as follows: in Section 2 we provide the basic details about 
observations and data reduction. Sections 3 and 4 concern the analysis of the X-ray images 
and spectra, respectively. The implications of our results on the nature of the halo are 
discussed in Section 5, and conclusion are drawn in Section 6. The luminosity distance of 107 
Mpc and the angular scale of 492~pc arcsec$^{-1}$ are assumed throughout for NGC~6240, 
based on the latest concordance cosmological parameters ($H_0=70.5$ km s$^{-1}$ 
Mpc$^{-1}$, $\Omega_m=0.27$, $\Omega_\Lambda=0.73$; Komatsu et al. 2011).

\section{Observations and Data Reduction}
This work is motivated by the most recent \textit{Chandra} observation of NGC~6240 (ObsID 
12713; PI: G. Fabbiano), which started on 2011 May 31 and yielded the deepest exposure of 
the source available to date ($\sim$145 ks). NGC~6240 had been previously targeted within 
the \textit{Chandra} guaranteed observing time program on several occasions. The imaging 
snapshot taken in February 2000 with the High Resolution Camera revealed the highly complex 
morphological properties in the central region, with the remarkable correlation between the 
X-ray emission contours and the H$\alpha$ filamentary structures (Lira et al. 2002). The 
following 37-ks observation (hereafter ObsID 1590) performed with the Advanced CCD 
Imaging Spectrometer (ACIS-S) detector clearly established the AGN nature of the two nuclei, 
which are both characterized by flat reflection spectra and prominent iron K$\alpha$ lines, 
and definitely confirmed the tight connection between the circumnuclear soft X-ray emission 
and starburst activity (Komossa et al. 2003). ACIS-S was also used in combination with the 
High-Energy Transmission Grating (HETG) in two consecutive observations in May 2006 
(ObsID 6908/9), for an integration time of $\sim$300 ks. The first order spectrum is quite 
faint (Shu et al. 2011), but results in a better resolution of the complex iron K-shell emission 
at 6.4--7.1 keV with respect to previous \textit{XMM-Newton}-based studies (Boller et al. 
2003; Netzer et al. 2005). 

The ACIS-S data products for ObsID 12713 provided by the \textit{Chandra} X-ray Center, as 
well as all the archival ones, were reprocessed through the \texttt{chandra\underline{ }repro} 
script using the v4.4.6 Calibration Database (CALDB), and analyzed with the \textsc{ciao} v4.4 
and \textsc{heasoft} v6.12 software packages. In the preliminary stages we also considered 
the two grating observations. However, differently from the spatially-resolved analysis of the 
nuclear hard X-ray emission presented in Wang et al. (2013a), the HETG zeroth orders turned 
out to be unsuitable for the study of soft, extended emission in the halo. Indeed, 
the 0.5--1.5 keV background level of ObsID 6908/9 is nearly 5 times higher than that of 
ObsID 12713. Combined with the large difference in effective area (by roughly a factor of 12 
around 1 keV),\footnote{\url{http://cxc.harvard.edu/cgi-bin/build_viewer.cgi?ea}.} this leads 
to a significant worsening of the overall signal-to-noise ratio (S/N) when all data sets are 
taken into account, against a modest improvement in terms of cumulative net counts (less 
than 10\%; Table~\ref{t1}). In addition, the soft X-ray emission clearly stretches out to 
the area of the detector over which the diffracted spectral orders are projected, preventing 
an accurate determination of the halo surface brightness along the HETG dispersion 
cones.\footnote{As a further caveat to the use of grating observations in the study of the 
halo, note that the estimated number of source counts for ObsID 6908/9 does not obviously 
scale with the exposure time (see Table~\ref{t1}).}

We therefore dropped ObsID 6908/9, and employed in this work the two standard ACIS-S 
observations only. In both cases the source was imaged on the back-illuminated S3 chip, due 
to its enhanced soft response. The background was low and fairly stable, with average rates 
of 0.0028 and 0.0039 counts s$^{-1}$ arcmin$^{-2}$ for ObsID 12713 and ObsID 1590; 
adopting a time bin of 300 s, the fit of the background light curves with a constant gives a 
reduced $\chi^2$ of 0.96 and 1.09, respectively. No time filtering of flaring background 
periods was then required, delivering a total exposure of 182.05 ks. By checking the field of 
view for X-ray counterparts of the optical point sources in the NOMAD catalogue (Zacharias 
et al. 2004), and comparing the positions of all the bright point-like X-ray sources in the two 
observations, we verified the accuracy of the absolute astrometry and the absence of any 
relative offset to within $\sim$1 native pixel (0.5$\arcsec$). The shorter-exposure ObsID 
1590 was then reprojected to the coordinate frame of ObsID 12713, and the two images 
were merged.\footnote{The choice of a specific script is not influential, hence we used 
\texttt{reproject\underline{ }events} plus \texttt{dmmerge} for the event files, and 
\texttt{reproject\underline{ }image} later on for the exposure maps.}

\begin{table}
\caption{\textit{Chandra} ACIS-S Observation Log for NGC~6240.}
\label{t1}
\begin{tabular}{c@{\hspace{25pt}}cccccc}
\hline \hline
ObsID$^*$ & Date & Exp$^a$ & Grating & $A_\rmn{eff}{}^b$ & Bkgd$^c$ & Counts$^d$ \\
\hline
1590 & 2001-07-29 & 36.7 & None & 630 & 0.0039 & 1960 \\
6909 & 2006-05-11 & 141.2  & HETG & 45 & 0.0014 & 430 \\
6908 & 2006-05-16 & 157.0 & HETG & 45 & 0.0015 & 350 \\
12713 & 2011-05-31 & 145.4 & None & 520 & 0.0028 & 5750 \\ 
\hline
\end{tabular}
\flushleft
\small{\textit{Note.} $^*$Grating observations were discarded for this work. $^a$Net 
exposure in ks; $^b$Nominal effective area at 1 keV in cm$^2$; $^c$Background count rate 
at 0.5--1.5 keV in s$^{-1}$ arcmin$^{-2}$; $^d$Estimated 0.5--1.5 keV source counts over 
the $r=20$--80$\arcsec$ region.}
\end{table}

Since the shape and the spatial extent of the diffuse emission are not known in advance, the 
regions for spectral extraction were optimized on the basis of the image inspection described 
in the next Section. Background spectra and count rates have been obtained from a circular 
region with radius of 1$\arcmin$, centered at a distance of $\sim$3$\arcmin$ from the 
nuclear region of NGC~6240, to the North-East, and devoid of both point sources and excess 
diffuse emission. Source and background spectra and weighted response files were generated 
and combined with the \texttt{specextract} tool. 

The X-ray halo is rather soft, and lies above the background level only in the $\sim$0.5--1.5 
keV energy range (which serves as reference throughout this work). It is therefore critical to 
adopt a correct spectral binning method. In order to allow the use of $\chi^2$ minimization, 
we have grouped the background-subtracted spectra to a significance threshold of 4$\sigma$ 
for each energy channel. At the same time, we have also employed the \textit{ungrouped} 
spectra, with the requirement of at least one count per bin to ensure a proper application of 
the Poissonian $C$-statistic (Cash 1979) even if no background model is 
supplied.\footnote{Since the background cannot be described by a simple shape (e.g., a 
power law), we did not conform to the ideal procedure with Poissonian statistics of modeling 
the total and background spectra simultaneously. However, it is still possible to obtain correct 
results within \textsc{xspec}, provided that no empty bin is involved.} The results of this dual 
approach are always self-consistent, and are presented interchangeably in the following, 
depending on the 
specific discussion purposes. Indeed, while $\chi^2$ carries direct information on the 
goodness of the fit, $C$-stat allows us to preserve a better energy resolution at lower 
S/N. The spectral analysis was performed with the \textsc{xspec} v12.7 fitting 
package.\footnote{\url{http://heasarc.gsfc.nasa.gov/xanadu/xspec/}.} Unless 
otherwise stated, all the uncertainties are given at the 90\% confidence level, and 
energies in the observer's frame. Radial profiles have been analyzed and fitted within 
\textit{Sherpa},\footnote{\url{http://cxc.harvard.edu/sherpa4.4/}.} while images 
were visualized and manipulated through SAOImage 
DS9.\footnote{\url{http://hea-www.harvard.edu/RD/ds9/}.}

\section{Image Analysis}
\subsection{Halo Morphology}
Besides the ordinary image reprocessing, further corrections are necessary in the study of 
extended X-ray sources to convert counts into physical units, properly taking into account 
the dependence of the collecting area on energy and position, as well as the effective 
exposure in the different regions of the detector. With this aim, we have first built an 
instrument map using the \textsc{ciao} tool \texttt{mkinstmap}, and then generated an 
exposure map (see Davis 2001) for each observation with \texttt{mkexpmap}. The input 
spectral weights were calculated by extracting a preliminary halo spectrum from the 
$r = 20$--80$\arcsec$ annulus centered on the brightest, nuclear hard X-ray source, and 
modeling it with a phenomenological two-component thermal model. The individual maps 
were then reprojected and combined to match the merged image. 

\begin{figure}
\includegraphics[width=15cm]{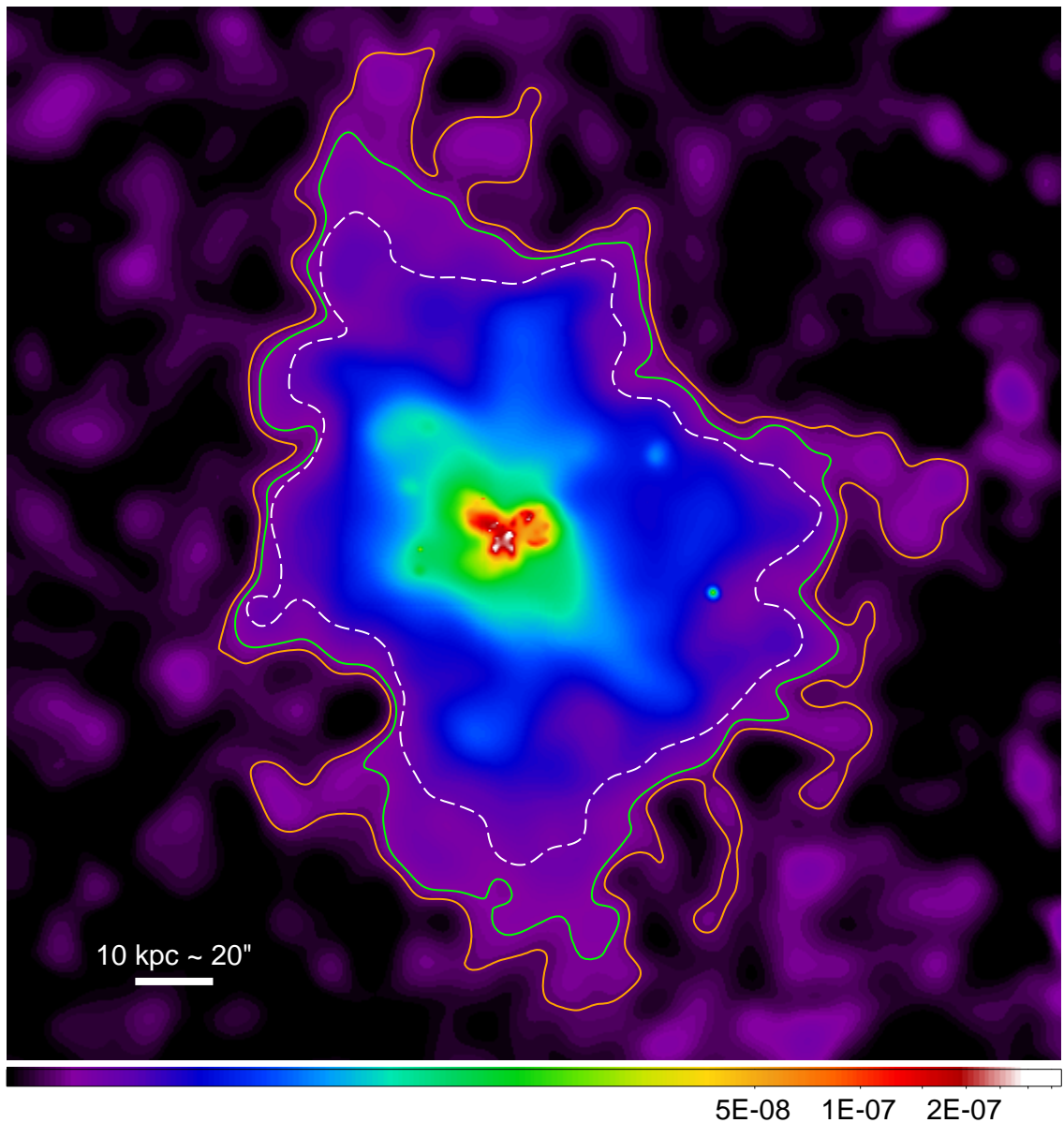}
\caption{Exposure-corrected, adaptively-smoothed halo image over the 0.7--1.1 keV 
energy range (North is up and East to the left). All the surrounding point sources have 
been removed. The orange and green solid curves represent the connected $2\sigma$ 
and $3\sigma$ detection contours above the average background level, which has been 
subtracted for clarity. For comparison, the white dashed curve is the $3\sigma$ 
confidence contour obtained from the 0.3--2 keV image, which is less sensitive to low 
surface brightness due to its higher background.} 
\label{hd}
\end{figure}
\begin{figure}
\includegraphics[width=15cm]{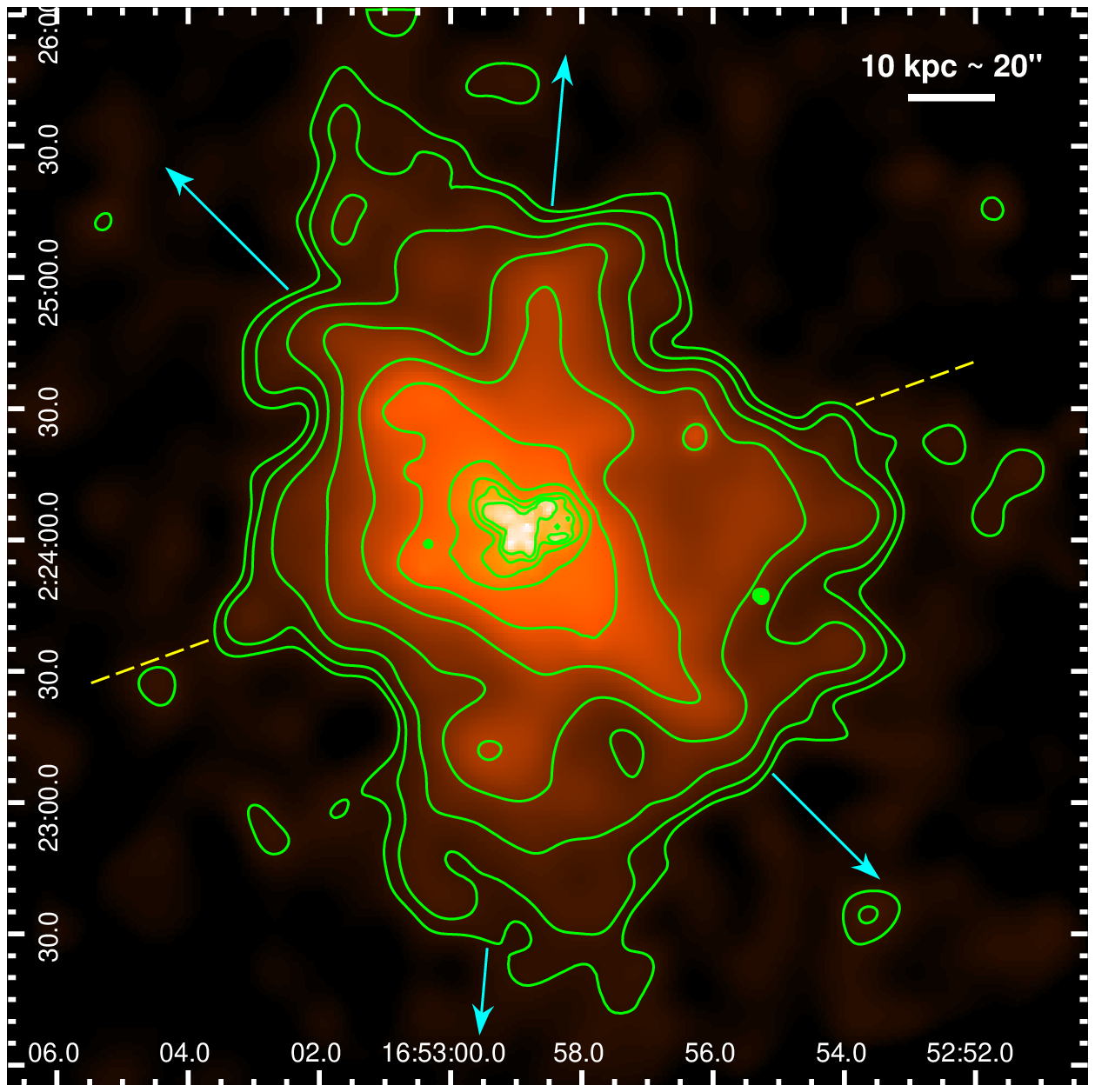}
\caption{Contours of the smoothed 0.7--1.1 keV emission, starting at the $3\sigma$ level 
above the background and increasing in logarithmic scale up to roughly one sixth of the 
maximum nuclear surface brightness, in order to emphasize the main structures in the halo. 
The four arrows trace the wide bicone generated by the prominent cross-like feature in the 
central $r \sim 1 \arcmin$, whose axis (dashed line) is virtually normal to the optical galactic 
disk.}
\label{hc}
\end{figure}

In order to assess the maximum halo extent along the various directions, we initially focused 
our attention on the 0.7--1.1 keV band, which represents the optimal compromise between 
the requirements of both high S/N and total net counts. At the standard soft X-ray energies 
of 0.3--2 keV, in fact, the source counts amount to $\sim$10000, but this is only 59\% of 
the whole number of events collected over that range. By restricting to 0.7--1.1 keV, instead, 
the same fraction rises to 77\% while retaining $\sim$6000 net counts. This improves the 
sensitivity at low surface brightness (see Fig.~\ref{hd}). 

In ObsID 12713, the nuclear region of NGC~6240 lies $\sim$2$\arcmin$ from the western 
edge of the S3 chip. Even after applying the exposure correction, this puts a rough empirical 
limit on the area that can be effectively explored for the presence of diffuse emission (at least 
in the East--West direction), as the large exposure disparity ($\approx 182/37$) would affect 
the significance of any detection. This notwithstanding, we have adaptively smoothed the 
0.7--1.1 keV image to emphasize the faint extended features while preserving the local details 
with the highest S/N, using \texttt{csmooth} (Ebeling et al. 2006). This tool convolves the raw 
image with increasing Gaussian kernels; we set a nominal significance between 2.5$\sigma$ 
and 5$\sigma$ over the local background under each kernel, and imposed a maximum kernel 
size of 10 pixels to prevent any oversmoothing. The same smoothing scales were applied to 
the exposure map, which was eventually used to normalize the source image. 

The final, exposure-corrected image is shown in Fig.~\ref{hd}, where also the detection contours 
at the 2$\sigma$ and 3$\sigma$ confidence level above the average background have been 
computed. The latter curve is taken to represent the outer boundary hereafter, although 
its mean radius of $\sim$80$\arcsec$ can still be regarded as conservative measure with 
some respect (e.g., the overly fine spatial resolution; see below). According to this estimate, 
the X-ray halo of NGC~6240 appears to have a somewhat distorted, elongated, diamond-like 
shape. The directions of minimum and maximum extension intersect to the South-East of the 
nucleus and are almost exactly perpendicular, corresponding to position angles of 
$\sim$16$\degr$ and 107$\degr$, respectively. The physical size at full length is then 
$\sim$110$\times$80 kpc 
(225$\arcsec \times$162$\arcsec$). In Fig.~\ref{hc} we draw the logarithmic contours 
starting at the fiducial detection limit, with a surface brightness cutoff that brings out the 
structures within the halo. As already pointed out by Bush et al. (2008), the most conspicuous 
feature is the stretched, cross-like pattern characterizing the central $r \sim 1 \arcmin$ region, 
whose arms generate a bicone with an aperture angle of $\sim$130$\degr$. This is even 
more evident below in Fig.~\ref{as}, where the emission clumpiness is visibly enhanced. 
Notably, the axis of symmetry of this bicone is virtually normal to the plane of the optical 
disk, and its apex is coincident with the hard X-ray source. This is all commensurate with a 
galactic outflow arising from the nuclear starburst similar to that observed in M82 (Griffiths 
et al. 2000), and possibly sweeping across a pre-existing halo medium. The physical scales 
involved, however, are rather challenging (see Section 5). 

\subsection{Surface Brightness Profile}
Compared to the oblong structure above, at larger distance ($r > 1 \arcmin \sim 30$ kpc) 
and out to the detection boundary the hot gas distribution tends to recover a rounder shape. 
In spite of the overall slight asymmetry, a radial analysis can provide valuable information 
on the system. We have then computed the surface brightness profile of NGC~6240 up to 
a galactocentric distance of 128$\arcsec$, sampling this region with successive, 
4$\arcsec$-wide annuli. The surrounding field is quite crowded, hence we employed the 
wavelet-based algorithm \texttt{wavdetect} (Freeman et al. 2002) to isolate any contaminating 
point-like sources, three of which lie just within the outer border. As their non-negligible 
contribution to the respective annuli might alter substantially the slope of the faint emission 
tail and the detection significance, we excluded a circle of 4-pixel radius around each of these 
sources. 

\begin{figure}
\includegraphics[width=15cm]{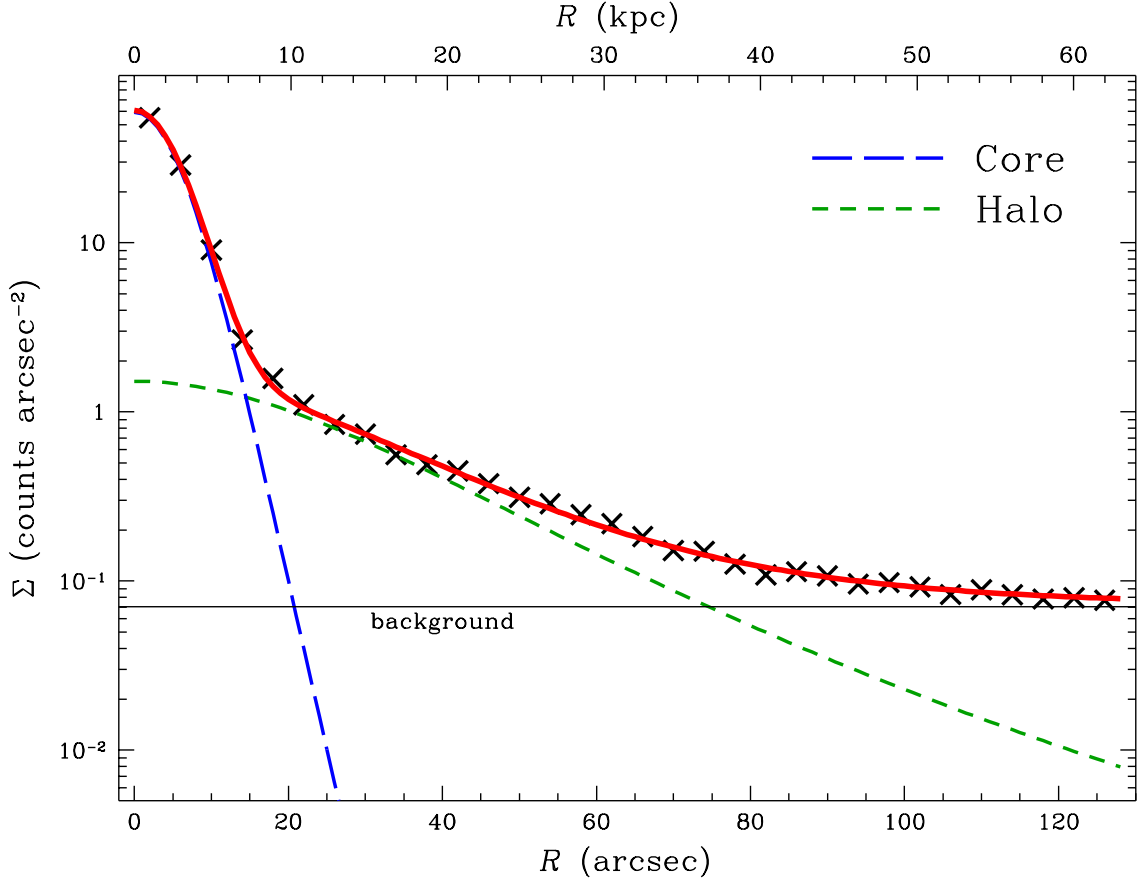}
\caption{Radial profile of the exposure-corrected surface brightness of NGC~6240 in the 
0.7--1.1 keV band. After the background subtraction, the profile is well described 
($\chi^2_\nu \sim 1.3$) by a double $\beta$-model (red solid curve), with a clear separation 
in the diffuse soft X-ray emission between the compact (\textit{core}) and the extended 
(\textit{halo}) component. Both are shown for clarity, respectively as the blue and the green 
dashed curve, while error bars are much smaller than the symbols and have been omitted. 
The background level (black thin line) is 0.07 counts arcsec$^{-2}$.}
\label{sb}
\end{figure}

The resulting background-subtracted surface brightness profile was fitted with a 
two-component $\beta$-model, which is a convenient analytical expression widely-used in 
the study of galaxies and clusters, defined as: 
\[
\Sigma(r)=\Sigma_0 \left[ 1+\left( \frac{r}{r_\rmn{s}} \right)^2 \right]^{-3\beta+0.5},
\]
where $\Sigma_0$ is the central surface brightness, $r_\rmn{s}$ the scale radius of the gas 
distribution and $\beta$ the power-law index. With $\beta=0.5$, for instance, $\Sigma 
\propto r^{-2}$ at $r \gg r_\rmn{s}$. Based on the shorter ObsID 1590 only, Huo et al. (2004) 
already suggested that a double $\beta$-model better accounts for the radial profile of 
NGC~6240 than the simpler form above. Besides the confirmation of these earlier findings, 
we are now able to show that the core and halo components can be completely disentangled 
(Table~\ref{t2}; Fig.~\ref{sb}). In particular, the extended, slowly declining component equals 
its compact, steeply falling counterpart $\sim$7 kpc away from the center, and becomes 
dominant straight beyond. We have consequently assumed $r = 15\arcsec$ as the inner 
boundary of the halo in our later spectral analysis. This is consistent with a rough estimate from 
visual inspection of the X-ray morphology. 

According to the radial surface brightness profile, diffuse emission associated with NGC~6240 
is detected at the 3$\sigma$ confidence level out to a galactocentric distance of 
$\sim$100$\arcsec$ (50 kpc). Since counts are now integrated over a given annulus, this 
measure naturally overrides the single pixel-based 3$\sigma$ contour formerly supplied, 
and is our best approximation to the actual average extension of the halo. 

\begin{table}
\caption{Best-fit Parameters of the Radial Surface Brightness Profile.}
\label{t2}
\begin{tabular}{c@{\hspace{25pt}}cc}
\hline \hline
Par & Core$^*$ & Halo$^*$ \\
\hline 
$\Sigma_0$ & 59.2$\pm$1.3 & 1.5$\pm$0.1 \\
$r_\rmn{s}$ & 10.0$\pm$1.5 & 23.8$\pm$2.5 \\
$\beta$ & 3.3$\pm$0.8 & 1.0$\pm$0.1 \\
\hline
\end{tabular}
\flushleft
\small{\textit{Note.} $^*$Component of the double $\beta$-model (errors are reported at 
$1\sigma$). $\Sigma_0$: central surface brightness in counts arcsec$^{-2}$; $r_\rmn{s}$: 
scale radius in kpc; $\beta$: power-law index.}
\end{table}

\subsection{Small-Scale Structures}
\begin{figure}
\includegraphics[width=15cm]{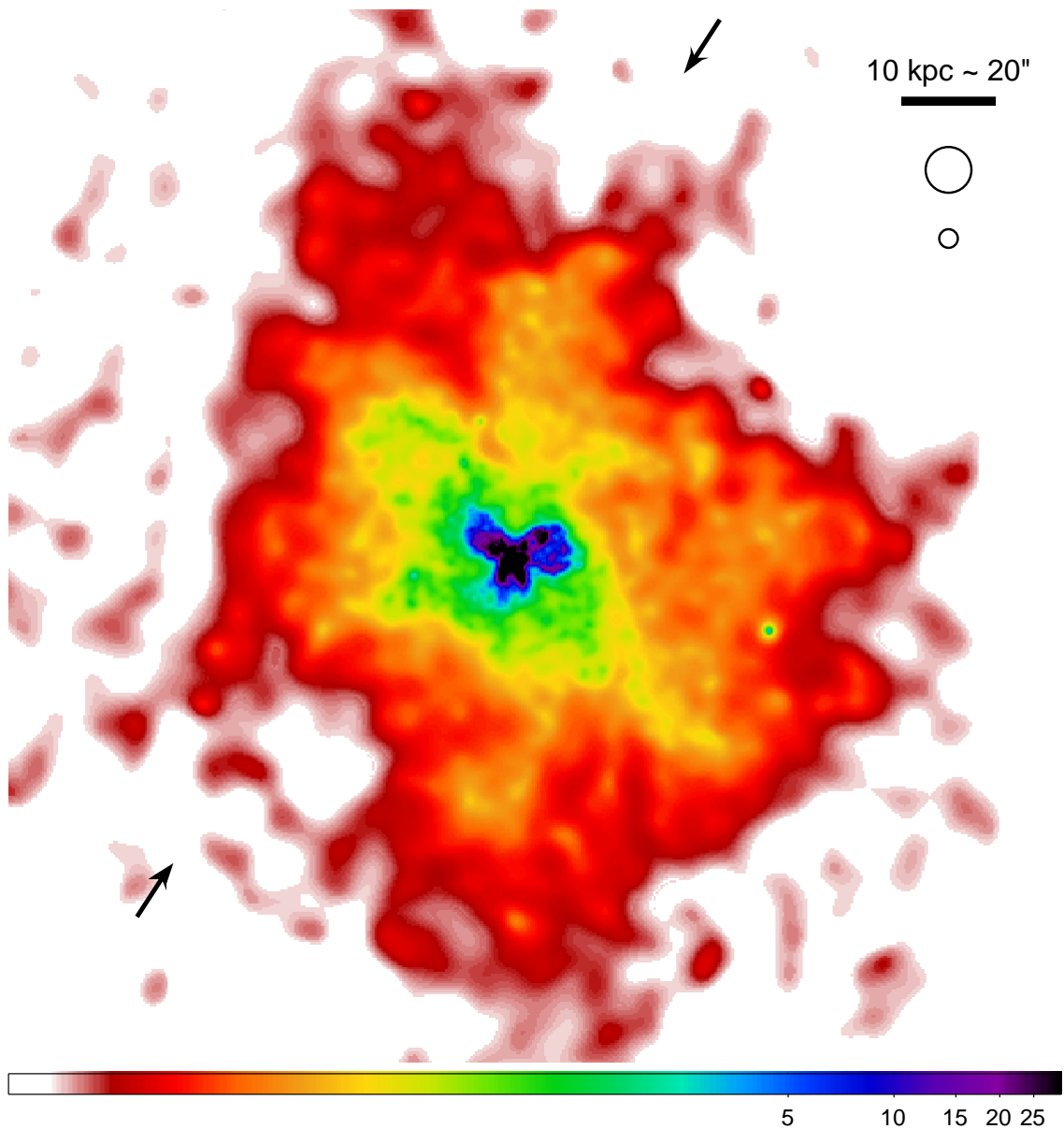}
\caption{Filamentary and clumpy nature of the extended soft X-ray emission in 
NGC~6240. The halo image in the 0.5--1.5 keV band has been adaptively smoothed with an 
algorithm that better preserves the spatial resolution of local structures when compared to 
Fig.~\ref{hd} (see the text for details). The larger circle in the upper right-hand corner 
corresponds to the maximum smoothing scale adopted, while the smaller one has a radius 
of 2$\arcsec$: structures of this size are significant up to the 3.5$\sigma$ level. The pair of 
arrows marks the position of the 200$\arcsec \times$4$\arcsec$ stripe whose close 
inspection is presented in Fig.~\ref{sf}.} 
\label{as}
\end{figure}

The clumpiness of the morphological appearance suggested by Fig.~\ref{hd}, as opposed to 
a possible flatness, can be itself indicative of the nature of the halo. In order to assess the 
degree of complexity, we have performed the adaptive smoothing with the alternative 
\textsc{ciao} tool \texttt{dmimgadapt}, which is also available within 
DS9.\footnote{\url{http://cxc.harvard.edu/ciao/ahelp/dax.html}.} Differently from 
\texttt{csmooth}, in this case the underlying algorithm returns a superior 
resolution of the bright small-scale structures. Indeed, within \texttt{csmooth} the kernel 
significance is computed with respect to the local background. This magnifies the faint, 
diffuse component close to the halo border, yet inner structures may be not significant 
enough with respect to the neighboring area. Such features are then smoothed at the widest 
scale allowed, and any clumpiness is lost. On the contrary, \texttt{dmimgadapt} is based on 
the raw number of counts. We then switched to the 0.5--1.5 keV image, due to its $\sim$13000 
total counts, 71\% of which are associated with the source. We adopted the same maximum 
size as before (10 pixels), and required at least 16 counts under each kernel. Given the 
background expected, this corresponds to a minimum significance of 1$\sigma$, growing up 
to 3.5$\sigma$ for the finest (i.e., $< 4\arcsec$-wide) structures. The results are illustrated 
in Fig.~\ref{as}, where knots and filaments are clearly visible throughout the physical scales 
across the halo.

\begin{figure}
\includegraphics[width=15cm]{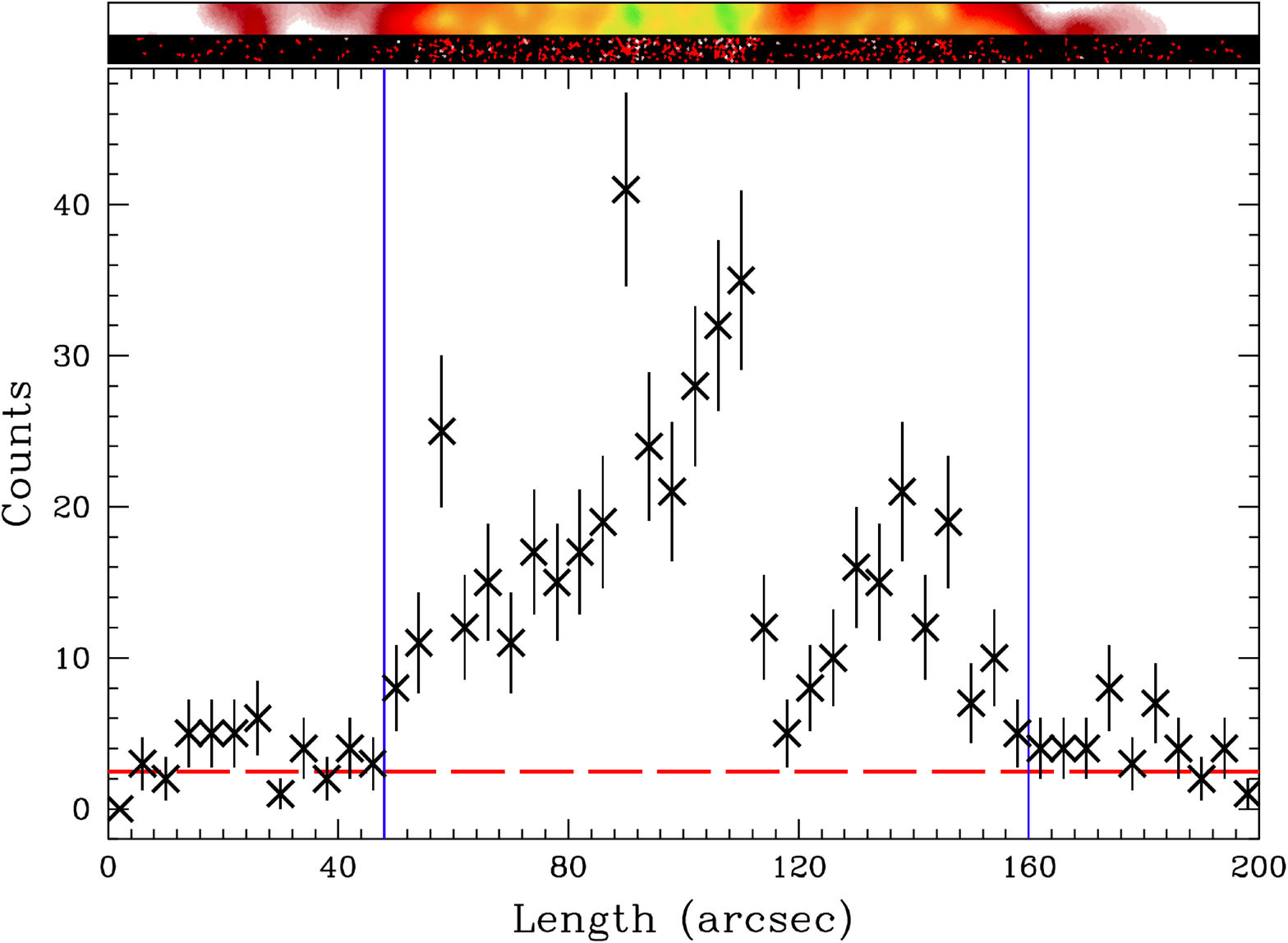}
\caption{Number of counts within the sequence of 50 square cells along the sample direction 
across the halo defined by the pair of arrows in Fig.~\ref{as}. The statistical significance of 
the small-scale structures revealed through the adaptive smoothing is definitely confirmed. 
The dashed red line represents the background expected over the cell size, while the vertical 
blue lines indicate the visual borders of the halo. The raw and smoothed images of the entire 
200$\arcsec \times$4$\arcsec$ stripe are reproduced for reference in the upper band.}
\label{sf}
\end{figure}

We further addressed this issue from a quantitative point of view by selecting a sample 
direction across the halo, off the nuclear region but intersecting several apparent, local 
intensity peaks. We extracted the number of counts along a 200$\arcsec \times$4$\arcsec$ 
stripe, divided in 50 square cells. Fig.~\ref{sf} shows the tight correlation between the raw 
and the smoothed image; the actual difference of count density substantiates the statistical 
significance of the small-scale structures. In conclusion, these features are not consistent 
with a fully relaxed environment in nearly thermal and hydrostatic equilibrium, but rather 
they are reminiscent of the loops, bubbles and cavities seen in the central $r < 7$ kpc region 
of NGC~6240, where the shock-heated gas is subject to violent starburst-driven winds and, 
possibly, to massive AGN feedback (Lira et al. 2002; Komossa et al. 2003; Feruglio et al. 
2012). A spatially-resolved spectral analysis can then provide crucial information to constrain 
the properties of the hot gas in the halo, and to shed light on the origin and evolution of the 
whole system.


\subsection{Line-Strength Map}
The preliminary spectrum extracted during the calculation of the exposure map 
hinted at soft X-ray line emission within the halo. Using the \textit{XMM-Newton} spectrum 
(70$\arcsec$ extraction radius) presented by Boller et al. (2003) as a reference, we were 
able to identify the Fe-L complex, possibly blended with O~\textsc{vii}--\textsc{viii} and 
Ne~\textsc{ix}--\textsc{x} (0.55--1.12 keV), and also Mg~\textsc{xi}--\textsc{xii} 
(1.24--1.5 keV) and Si~\textsc{xiii}--\textsc{xiv} lines (1.7--1.93 keV). Following Baldi et al. 
(2006a) and Wang et al. (2009), we generated three narrow-band images to encompass the 
features listed above, in order to obtain a visual map of line strengths and metal abundances 
in the hot, diffuse interstellar medium. For the continuum subtraction, another image was 
extracted from the joint 0.4--0.55, 1.12--1.24 and 1.5--1.7 keV energy range. All the line 
and continuum bands were selected to achieve an optimal compromise between line strength 
and continuum contamination. No culling was applied for the few point sources. The 
O--Fe--Ne image was adaptively smoothed (with \texttt{csmooth}) as described earlier. Due 
to its highest S/N and largest number of line counts ($\sim$4000), the same smoothing 
scales were adopted for the other images. The background level appropriate to each energy 
range was then subtracted, while the continuum contribution to each line image was 
determined by fitting the halo spectrum with an absorbed bremsstrahlung model, after 
excluding the strong emission features. The resulting scaling factors ($\sim$2.2, 0.4 and 
0.1 for the O--Fe--Ne, Mg and Si bands, respectively) were assumed as relative weights in 
the final continuum subtraction. 

\begin{figure}
\includegraphics[width=\textwidth]{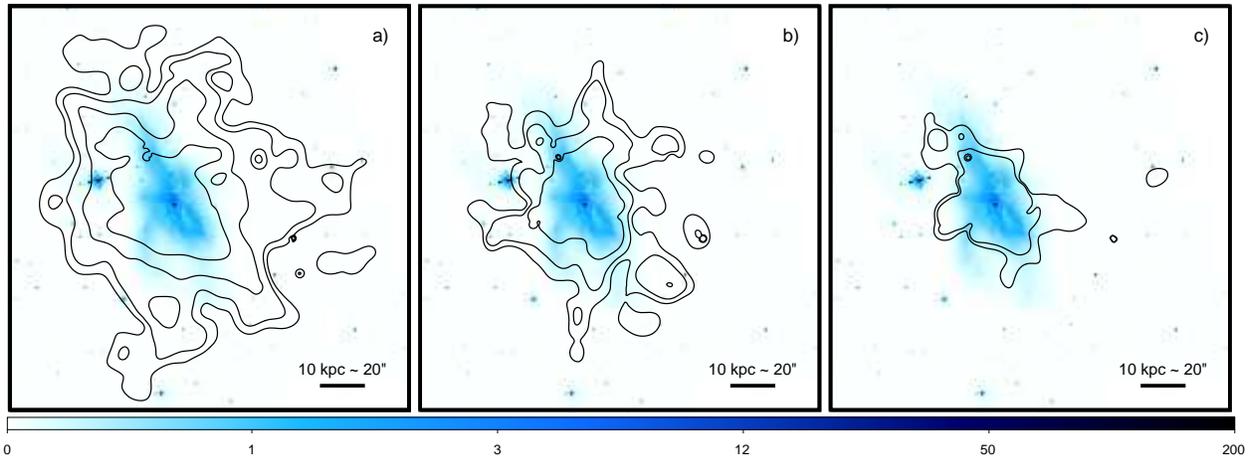}
\caption{Contours of the soft X-ray emission lines superimposed on the \textit{Hubble Space 
Telescope} ACS/WFC F814W image of NGC~6240 (Cycle 14, Program 10592; PI: A. Evans. 
Retrieved from the NASA Multimission Archive at the Space Telescope Science Institute). From 
left to right: a) 3$\sigma$, 5$\sigma$, 10$\sigma$ and 20$\sigma$ of the O--Fe--Ne blend 
at 0.55--1.12 keV; b) 3$\sigma$, 5$\sigma$ and 10$\sigma$ of Mg~\textsc{xi}--\textsc{xii} 
at 1.24--1.5 keV; c) 3$\sigma$ and 5$\sigma$ of Si~\textsc{xiii}--\textsc{xiv} at 1.7--1.93 keV.}
\label{lm}
\end{figure}

The accuracy of this procedure is corroborated by the little spectral variations across the 
halo (see below), which allowed us to employ the average continuum. Indeed, this is not 
the case for the core region, where the source emission is much more complex and 
position-dependent. As a result, most of the structures found in the line-strength maps 
within the central 15$\arcsec$ have no physical meaning. For this reason, in Fig.~\ref{lm} 
we only show the lowest significance contours of the soft X-ray emission lines under 
examination. The comparison with the archival \textit{Hubble Space Telescope} ACS/WFC 
F814W image reveals that metals are spread out far beyond the optical limits of the galaxy. 
Not surprisingly, given the energy band and the line intensity, the O--Fe--Ne morphology 
turns out to be very similar to that illustrated in Fig.~\ref{hc}. Once the intrinsic difference 
in S/N is taken into account, also the spatial distribution of Mg and Si line emission displays 
a close overlap. In particular, the prominent cross-like shape is clearly recovered in the 
3$\sigma$ Mg contours. This suggests that the halo of NGC~6240 has already experienced 
a significant, nearly uniform metal enrichment out to very large distances. 

\section{Spectral Analysis}
\subsection{Full Halo}
\begin{figure}
\includegraphics[width=15cm]{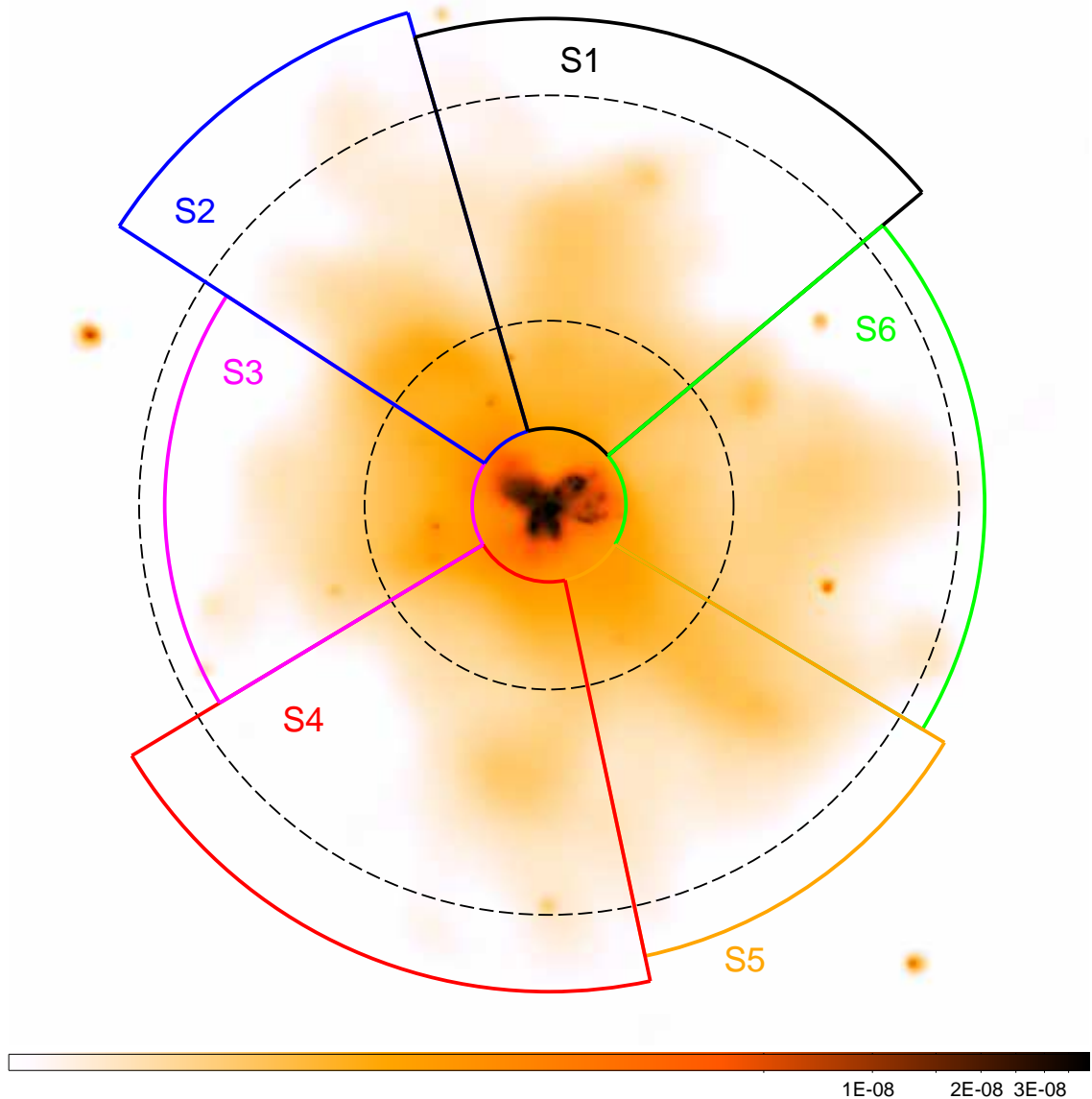}
\caption{Definition of the spectral extraction region for the full halo (FH), obtained as the 
compilation of six azimuthal sectors, S1--S6, which are later used for the spatially-resolved 
analysis. Their selection is based on both the total number of net counts and the morphological 
features, and follows qualitatively the 3$\sigma$ contour above the background in the smoothed 
0.5--1.5 keV halo image (whose level corresponds to the zero point of the color scale). The thin 
dashed lines refer to the inner (IH; $r = 15$--36$\arcsec$) and outer (OH; $r = 36$--80$\arcsec$) 
halo regions.} 
\label{hm}
\end{figure}

The radial surface brightness profile of Fig.~\ref{sb} indicates a sharp separation at 
$r = 15\arcsec \simeq 7.5$ kpc between the core and the halo components. Starting at this 
inner boundary, we followed the $3\sigma$ contours of the 0.5--1.5 keV image to define 
the width of the extraction region for the spectral analysis. Due to the slight asymmetry, the 
outer border is not fixed, but varies from 75$\arcsec$ to 100$\arcsec$ over six azimuthal 
sectors (labeled S1--S6; see Fig.~\ref{hm}). Qualitatively then, the full halo (FH) region has 
the shape of an irregular windmill, and contains $\sim$11300 net counts over the entire 
0.3--8 keV energy range. With no loss of statistical information, we restricted to 0.4--2.5 
keV ($\sim$10700 net counts) for the spectral fitting. 

Ten point-like, soft X-ray sources were identified within the halo extraction boundaries by 
running \texttt{wavdetect} over the selected band. Each of them yields a few tens of counts at 
most, for an aggregate contribution to the extended emission of less than 3\%. Given their minor 
weight, these point sources were not excluded. Their cumulative spectrum is well reproduced 
by an absorbed ($N_\rmn{H} \sim 10^{21}$ cm$^{-2}$) power law with photon index set to 
1.8, whose intrinsic 0.3--8 keV luminosity is $< 3 \times$10$^{40}$ erg s$^{-1}$ at the 
distance of NGC~6240. Owing to their location, six are allegedly background AGN (but no 
optical counterpart is known), while the remaining four, if local to the galaxy, would be 
definitely ultraluminous X-ray sources (ULXs; e.g., Zezas \& Fabbiano 2002). The observed 
spectral shape is also consistent with the unresolved populations of X-ray active objects, 
including high- and low-mass X-ray binaries, active binaries, cataclysmic variables (e.g., 
Fabbiano 2006; Boroson et al. 2011). These sources are associated with the stellar 
population, hence their contribution to the halo is expected to be negligible. We accounted 
for any residual emission from the latter classes (as well as for the resolved point sources) by 
keeping the $\Gamma = 1.8$ power-law component in our model. 

The bulk of the soft X-ray emission in the halo is presumed to arise from diffuse hot gas, 
and was then modeled as a thermal spectrum through the \texttt{vapec} code (Smith et al. 
2001), which makes use of the AtomDB v2.0.1 atomic 
database.\footnote{\url{http://atomdb.org}. An update to v2.0.2 has been recently 
released (Foster et al. 2012), but our results are not modified (best-fit abundances 
are only affected at the third digit).} The gas emission 
measure (EM) can be expressed as a function of the \texttt{vapec} normalization 
$\mathcal{N}_\rmn{v}$ through the relation $\rmn{EM}=\int n_e n_\rmn{H} 
\mathrm{d}V=4\pi[D_\rmn{A}(1+z)]^2 \times 10^{14} \mathcal{N}_\rmn{v}$, where $n_e$ 
and $n_\rmn{H}$ are the electron and hydrogen densities, and $D_\rmn{A}$ is the angular 
diameter distance to NGC~6240 (102 Mpc). We first considered a single-temperature plasma 
component, allowing for local (i.e., at the redshift of the source) absorption in addition to the 
Galactic column density ($N_\rmn{H} = 4.87 \times 10^{20}$ cm$^{-2}$; Kalberla et al. 
2005). The model form is then expressed within \textsc{xspec} as 
\texttt{wabs*zwabs*(vapec+powerlaw)}. As mentioned, the halo region provides 
$\sim$10700 counts, almost 90\% of which are detected at 0.5--1.5 keV,  with an 
improvement by a factor of $\sim$4 with respect to ObsID 1590 alone. This enabled us to 
assess for the first time individual abundances for iron and each of the main 
$\alpha$-elements (O, Ne, Mg, Si). For all the other elements we have adopted solar 
abundances (from Anders \& Grevesse 1989). Even if this standard 
practice might lead to some inaccuracies, it is not really critical here, as the overall contribution of the 
frozen elements to the best fit is marginal ($<$~20\%). The only possible exception is represented 
by nickel L-shell emission ($\sim$6\%; see later). We refer to Kim 2012 for a review of all the systematic 
uncertainties affecting the measure of elemental abundances through X-ray spectral fitting to CCD data. 

\begin{figure}
\includegraphics[width=12cm]{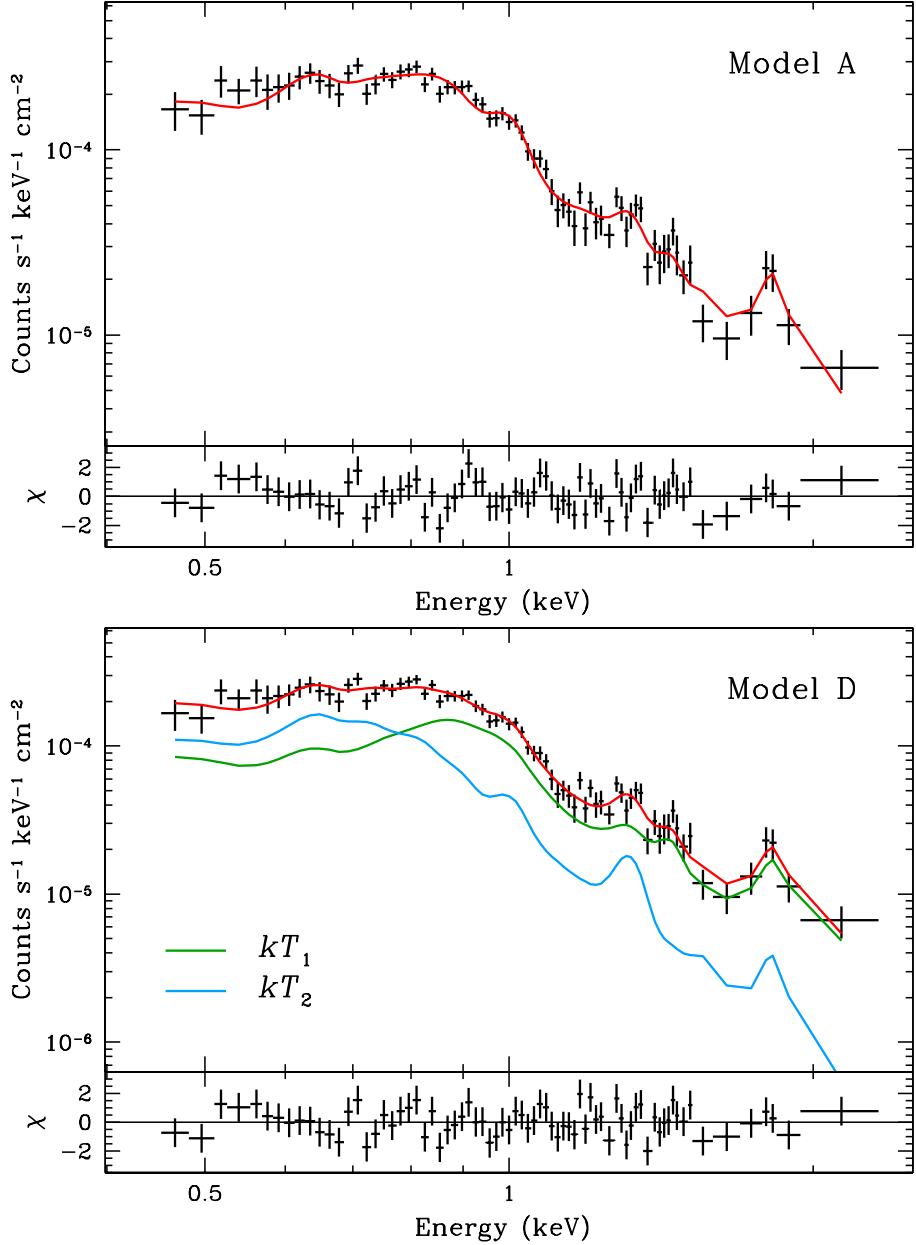}
\caption{\textit{Chandra} ACIS-S spectrum of the soft X-ray halo of NGC~6240 (FH region), rebinned to a 
4$\sigma$ significance for each energy channel and fitted with the $\chi^2$ statistic. Top panel: best fit 
and residuals (in units of $\sigma$) for the single-temperature hot gas emission (model A). Bottom panel: 
same as above, but allowing for a two-temperature plasma (model D). Both thermal components are shown, 
with $kT_1 \sim 0.9$ keV and $kT_2 \sim 0.4$ keV.} 
\label{hf}
\end{figure}

The spectrum of the halo region is plotted in Fig.~\ref{hf}, while the best-fitting parameters are 
listed in Table~\ref{t3}. Model A applies to the rebinned spectrum, for which the $\chi^2$ statistic 
is assumed, while model B is relative to the ungrouped data, treated with $C$-stat (see the Appendix 
for an overview). The results are in full agreement. Interestingly, the amplitude of the power-law 
continuum and the column density of the local absorber have a null best-fit value, implying that 
both components are not required, and that the diffuse, soft X-ray emission can thus be accounted 
for within a simple thermal scenario ($\chi^2_\nu \sim 1.14$). The upper limit to the 0.3--8 keV 
power-law luminosity is 4.1$\times$10$^{40}$ erg s$^{-1}$, encompassing the contribution 
estimated above for the point sources. On the other hand, NGC~6240 is known to host a huge 
amount of cold gas, but this is concentrated in the nuclear regions (Baan et al. 2007). In model C 
we fitted the six halo subregions S1--S6 simultaneously with a \texttt{wabs*vapec} model, where 
the thermal emission is modified by Galactic absorption only. Individual temperatures and metal 
abundances are tied, and are found to be in good agreement with the fiducial values. 

\begin{table}
\caption{Best-fitting Thermal Models of the Full Halo Region.} 
\label{t3}
\begin{tabular}{lccccc}
\hline \hline
Mod$^*$ & A & B & C & D & E \\
\hline 
$N_\rmn{H}(z)$ & $< 1.8$ & $< 1.6$ & ... & ... & ... \\
$f_\mathit{pl}$ & $< 0.06$ & $< 0.05$ & ... & ... & ... \\
$kT_1$ & 0.65$^{+0.06}_{-0.03}$ & 0.66$^{+0.06}_{-0.04}$ & 0.66$^{+0.06}_{-0.03}$ & 
0.89$^{+0.07}_{-0.23}$ &  0.77$^{+0.09}_{-0.07}$ \\
$kT_2$ & ... & ... & ... & 0.42$^{+0.11}_{-0.08}$ & 0.26$^{+0.06}_{-0.08}$ \\
EM$_1$ & 40.5$^{+8.0}_{-6.6}$ & 38.7$^{+7.3}_{-7.0}$ & 39.5$^{+5.6}_{-5.3}$ & 
20.1$^{+15.5}_{-4.9}$ & 28.1$^{+7.8}_{-7.1}$ \\
EM$_2$ & ... & ... & ... & 17.5$^{+6.2}_{-5.3}$ & 38.9$^{+14.4}_{-18.2}$ \\
$L_\mathit{th}$ & 3.76$^{+0.44}_{-0.98}$ & 3.73$^{+0.42}_{-0.84}$ & 3.74$^{+0.15}_{-0.16}$ & 
3.81$^{+0.40}_{-0.98}$ & 3.84$^{+0.63}_{-1.12}$ \\
$Z_\rmn{O}$ & 0.50$^{+0.33}_{-0.21}$ & 0.54$^{+0.39}_{-0.20}$ & 0.53$^{+0.22}_{-0.17}$ & 
0.44$^{+0.29}_{-0.17}$ & 0.22$^{+0.53}_{-0.22}$ \\
$Z_\rmn{Ne}$ & 0.57$^{+0.29}_{-0.21}$ & 0.62$^{+0.31}_{-0.20}$ & 0.58$^{+0.21}_{-0.17}$ & 
0.39$^{+0.27}_{-0.26}$ & 0.83$^{+0.63}_{-0.43}$ \\
$Z_\rmn{Mg}$ & 0.43$^{+0.18}_{-0.15}$ & 0.46$^{+0.20}_{-0.14}$ & 0.43$^{+0.14}_{-0.11}$ & 
0.68$^{+0.38}_{-0.28}$ & 0.76$^{+0.46}_{-0.29}$ \\
$Z_\rmn{Si}$ & 0.46$^{+0.21}_{-0.18}$ & 0.47$^{+0.21}_{-0.18}$ & 0.44$^{+0.15}_{-0.13}$ & 
0.59$^{+0.31}_{-0.24}$ & 0.57$^{+0.34}_{-0.23}$ \\
$Z_\rmn{Fe}$ & 0.11$^{+0.03}_{-0.02}$ & 0.12$^{+0.03}_{-0.02}$ & 0.11$^{+0.03}_{-0.01}$ & 
0.18$^{+0.08}_{-0.06}$ & 0.15$^{+0.08}_{-0.04}$ \\
$\chi^2/\nu$ & 74.2/65 & ... & ... & 66.1/65 & 63.4/65 \\
$C$-stat/$\nu$ & ... & 118/124 & 684/685 & ... & ... \\
\hline
\end{tabular}
\flushleft
\small{\textit{Note.} $^*$See the text and the Appendix for both the definition of the models 
and their different assumptions.  $N_\rmn{H}(z)$: local column density in 10$^{20}$ cm$^{-2}$; 
$f_\mathit{pl}$: relative contribution of the power-law component to the observed 0.4--2.5 keV 
emission; $kT$: plasma temperature in keV; EM: emission measure in 10$^{63}$ cm$^{-3}$; 
$L_\mathit{th}$: intrinsic 0.4--2.5 keV thermal luminosity in 10$^{41}$ erg s$^{-1}$; 
$Z_\rmn{X}$: elemental abundances in solar units.}
\end{table}

The main result is a general metal underabundance (by a factor of $\sim$2 for $\alpha$-elements) 
with respect to solar (Table~\ref{t3}). Absolute abundances should be taken with some caution, since 
they strongly depend on the determination of the continuum, which, in turn, is affected by unknown 
factors such as absorption, non-thermal components and multi-temperature gas emission (e.g., Kim 
2012). We then tried introducing a second thermal component to probe the possible presence in the 
halo of multi-temperature gas. Due to the heavy degeneracy between the column density and the 
temperature of the additional gas phase, which favors a statistically equivalent (but physically 
unacceptable) solution,\footnote{A column of $N_\rmn{H} \sim 3 \times 10^{21}$ cm$^{-2}$ over 
the full halo extent would imply a mass of cold gas of $\sim$10$^{11} M_{\sun}$.} we have dropped 
both local absorption and power-law continuum, adopting a \texttt{wabs*(vapec+vapec)} form 
(model D). The goodness of fit is significantly increased ($\chi^2_\nu \sim 1.02$), and calls for a 
mixed plasma with $kT_1 \sim 0.9$ keV and $kT_2 \sim 0.4$ keV, while the two emission measures 
are comparable. The results are illustrated in Fig.~\ref{hf}, and summarized in Table~\ref{t3}. Notably, 
there is no substantial change in the abundances of iron and of the main $\alpha$-elements, which 
were still allowed to be different fractions of solar, but were tied between the two gas components 
since they cannot be constrained separately. In model E, instead, abundance was fixed to 0.1 solar 
for all the elements in the low-temperature component. Both $kT_1$ and $kT_2$ decrease by 
$\sim$0.15 keV, yet the fit is further improved ($\Delta \chi^2 \simeq -2.7$ with the same degrees 
of freedom), supporting the presence of complex physical conditions in the halo of NGC~6240. This 
notwithstanding, the single-temperature model clearly represents a sound approximation to the data, 
and provides useful parameters for the purpose of this work. 

As a further test, we considered the possibility that collisional ionization equilibrium (the 
underlying assumption of the \texttt{vapec} code) is not met by the shock-heated gas in the halo.
NGC~6240 is undergoing a violent, merger-induced starburst phase, which is responsible for most 
of the system's $\sim$10$^{12} L_{\sun}$ IR luminosity (Lutz et al. 2003; Nardini et al. 2009). The 
last major burst of star formation is estimated to have occurred $\sim$20 Myr ago (Tecza et al. 
2000). Whether a pre-existing halo medium is swept up and shock-heated by a starburst-driven 
wind, or shocks are taking place inside the wind material itself, a low-density ($n_e \sim 10^{-3}$ 
cm$^{-3}$) gas may have not recovered ionization equilibrium conditions yet. The typical, 
density-weighted time-scale for plasma to reach equilibrium is 
$\tau_\rmn{eq} = n_e t > 10^{12}$ cm$^{-3}$ s (Masai 1994; Smith \& Hughes 2010). We then 
switched from \texttt{vapec} to the 
non-equilibrium model \texttt{vnei} (Borkowski et al. 2001), fixing the ionization time-scale 
parameter at $\tau_\rmn{eq} =10^{11}$ cm$^{-3}$ s. This returned a much poorer spectral 
description ($\chi^2_\nu \sim 2$), hence we did not pursue this interpretation any further. 
Incidentally, by taking $\tau_\rmn{eq} = 3 \times 10^{12}$ cm$^{-3}$ s we obtained 
$\chi^2_\nu \sim 1.3$, proving that \texttt{vnei}, if unconstrained, tends to converge towards 
an equilibrium solution.  

Finally, we investigated the nickel L-shell issue in more detail, by repeating the spectral fits 
(model A) with different values of nickel abundance. This was in turn fixed to half-solar, solar and 
twice-solar, tied to iron and left free. This check delivers a nearly evenly-spaced grid of $Z_\rmn{Ni}$ 
values, from which we can draw some qualitative considerations in spite of the large uncertainties 
involved. Although the corresponding best fit is still broadly acceptable (null hypothesis probability 
of $\sim$11\%), the assumption of $Z_\rmn{Ni} \equiv Z_\rmn{Fe} \simeq 0.15$ results in a 
significant worsening on statistical grounds ($\Delta \chi^2 \simeq 5.1$). Indeed, nearly solar values 
of $Z_\rmn{Ni}$ are definitely preferred. Equivalent results were found with $C$-stat on the ungrouped 
spectra. Once an obvious nickel underabundance is ruled out, most elements are virtually unaffected 
by the exact entry for $Z_\rmn{Ni}$, due to the substantial difference from nickel in either total 
contribution (Fe) or energy (O, Mg, Si). Conversely, neon abundance varies systematically with 
$Z_\rmn{Ni}$, decreasing by a factor of $\sim$3 as nickel changes from half to twice solar, as the 
two elements are very similar in both flux and energy. Thawing $Z_\rmn{Ni}$ has a negligible effect 
on the goodness of fit ($\Delta \chi^2 \simeq -0.6$). An $F$-test gives a probability of chance 
improvement of 48\%. The best-fit value is $Z_\rmn{Ni} \sim 1.4$, but poorly constrained in the 
0.5--2.5 range. We have therefore kept nickel abundance frozen to solar for all the subsequent 
analysis.

\subsection{Radial Analysis}
Given the fairly large amount of total net counts available, the next step is to embark upon a 
spatially-resolved analysis aimed at revealing any temperature and/or metallicity gradient within 
different zones of the halo. We first searched for some trend in the radial direction, and selected 
two annular regions corresponding to the inner (IH; $r = 15$--36$\arcsec$) and outer (OH; 
$r = 36$--80$\arcsec$) halo (see Fig.~\ref{hm}). The IH/OH transition radius was adaptively chosen to 
ensure an equal number of source counts ($\sim$4500) in the 0.5--1.5 keV energy band. Due 
to the different areas (by a factor of $\sim$5), however, the S/N in the two subregions is quite 
different (see Table~\ref{t4}). In the following, we systematically refer to the analysis of the 
ungrouped spectra, which preserve the intrinsic ACIS-S energy resolution. This entails sounder 
measures and more reliable comparisons among elemental abundances, as Mg and Si lines 
fall in a rather noisy spectral range. Even the O--Fe--Ne blend can be affected by a conservative 
rebinning like the one adopted to employ the $\chi^2$ minimization. 

\begin{table*}
\caption{Summary of the Different Extraction Regions.}
\label{t4}
\begin{tabular}{cccc@{\hspace{30pt}}cccc}
\hline \hline
Reg & $\mathcal{C}_{0.5-1.5~\rmn{keV}}$ & $\mathcal{C}_{0.4-2.5~\rmn{keV}}$ & $\mathcal{F}_{s/t}$ &
Reg & $\mathcal{C}_{0.5-1.5~\rmn{keV}}$ & $\mathcal{C}_{0.4-2.5~\rmn{keV}}$ & $\mathcal{F}_{s/t}$ \\
\hline
FH & 9463$\pm$147 & 10675$\pm$173 & 63.7 & IH & 4560$\pm$72 & 5176$\pm$79 & 86.2 \\
S1 & 1577$\pm$52 & 1742$\pm$60 & 58.2 & OH & 4526$\pm$103 & 5084$\pm$120 & 56.6 \\
S2 & 1575$\pm$48 & 1777$\pm$54 & 67.3 & H1 & 1837$\pm$44 & 2108$\pm$48 & 91.6 \\
S3 & 1583$\pm$47 & 1775$\pm$52 & 70.5 & H2 & 1802$\pm$45 & 2044$\pm$50 & 84.8 \\
S4 & 1579$\pm$53 & 1775$\pm$61 & 57.4 & H3 & 1826$\pm$48 & 2032$\pm$53 & 76.7 \\
S5 & 1581$\pm$47 & 1788$\pm$53 & 69.3 & H4 & 1816$\pm$52 & 2030$\pm$59 & 66.6 \\
S6 & 1568$\pm$50 & 1820$\pm$57 & 63.3 & H5 & 1804$\pm$70 & 2046$\pm$83 & 44.7 \\
\hline
\end{tabular}
\flushleft
\small{\textit{Note.} $\mathcal{C}$: net counts over the reference energy band (0.5--1.5 keV) and over 
the full spectral range (0.4--2.5 keV); $\mathcal{F}_{s/t}$: fraction of source to total counts at 0.4--2.5 keV, 
in per cent.}
\end{table*}

\begin{figure}
\includegraphics[width=15cm]{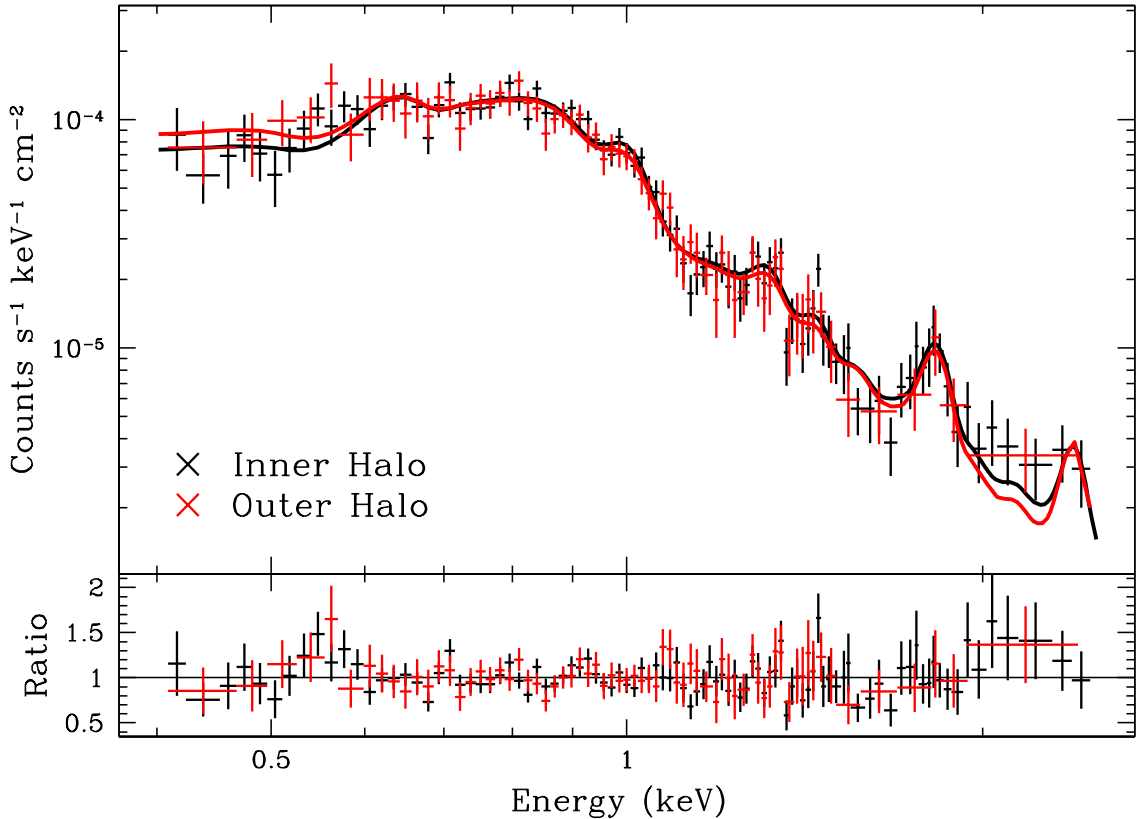}
\caption{Source spectra, best-fitting models and data/model ratios for the inner (IH, black) 
and outer (OH, red) halo regions. The data were rebinned for plotting purposes only, and no 
cross-scale factor is introduced.} 
\label{io}
\end{figure}

\begin{figure}
\includegraphics[width=15cm]{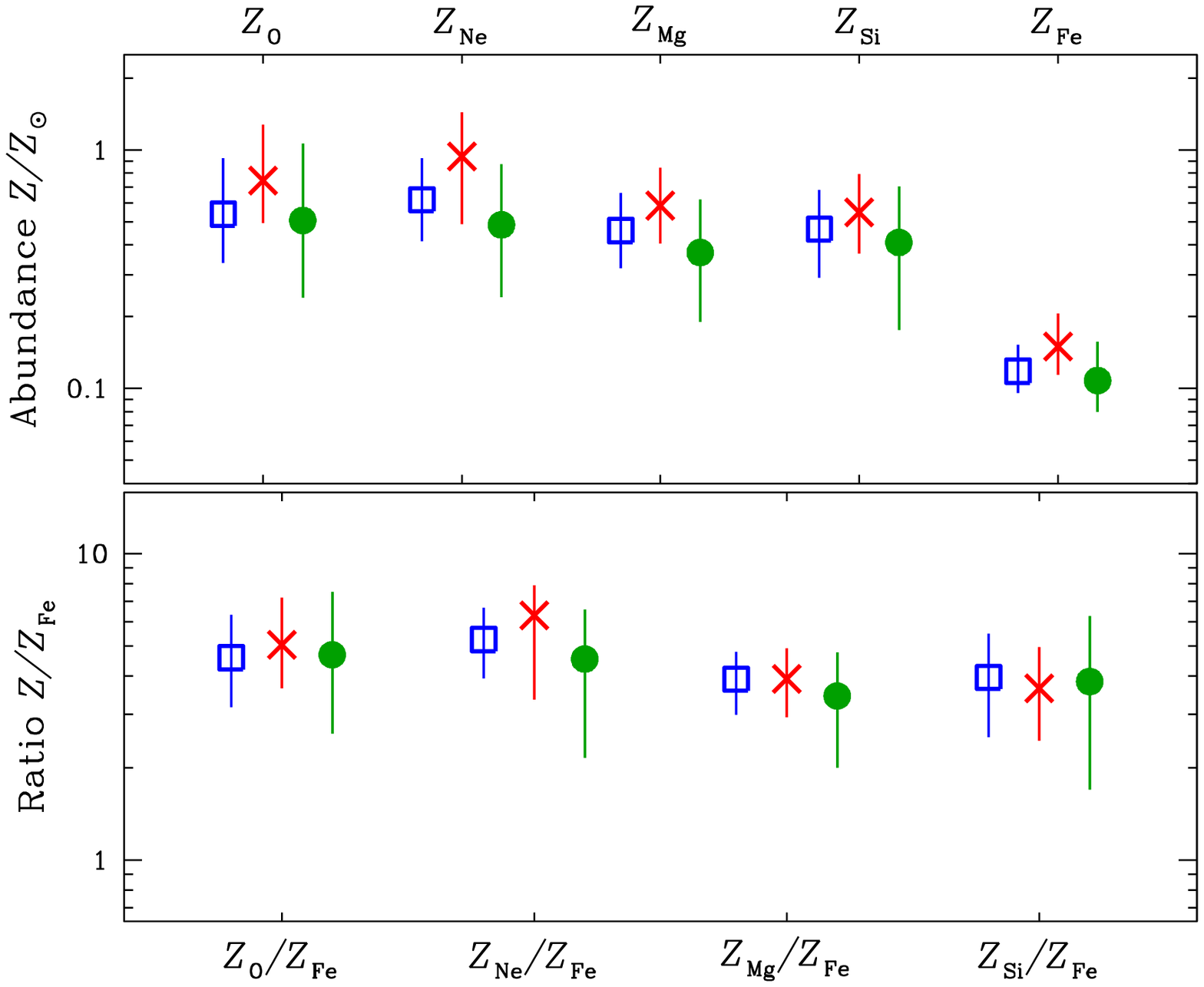}
\caption{Top panel: absolute abundances (in solar units) of all the main elements, as derived 
from the spectral analysis of the inner (red crosses) and outer (green dots) halo regions. 
Although the values for individual elements are consistent with each other, the two sets of 
measures hint at a possible decreasing trend of metallicity with distance. The cumulative 
significance is anyway limited to 2$\sigma$. The average values over the full halo are also 
plotted for comparison (blue squares). Bottom panel: relative abundances with respect to iron, 
showing no evident radial dependence.}
\label{rd}
\end{figure}

We have then used model B for the IH and OH fits. A marginal power-law component is possibly 
present in the IH region, consistent with the luminosity of the point sources that fall in that field. 
The upper limit to the total power-law contribution is in good agreement with the previous 
estimates in Table~\ref{t3}. The two IH and OH plasma temperatures are identical to the fiducial 
FH value, and again no local absorber is required. Ultimately, the two spectra are remarkably alike 
(Fig.~\ref{io}). The most significant comparison concerns abundances, as illustrated in Fig.~\ref{rd}. 
In spite of the overlapping error bars, for each element the best-fitting OH values are systematically 
lower than the IH ones, suggesting a slight metallicity drop at larger distances. By disentangling the 
core and halo spectra, a similar result had been already found by Huo et al. (2004), who used a 
simplified model to fit the ObsID 1590 data (see also Grimes et al. 2005). Although negligible 
when considering the single element, the overall effect is significant at the 2$\sigma$ level. As 
already mentioned, however, absolute abundances can be somewhat misleading. Relative 
abundances (i.e., abundance ratios between two elements) are usually more reliable, as the 
prominence of the different lines is altered in a similar way by any incorrect subtraction of the 
underlying emission. The decreasing trend is not seen in the abundance ratios. This can be 
explained in either of the following ways: \textit{(1)} the observed metallicity drop is artificial, 
likely induced by an inaccurate determination of the continuum; \textit{(2)} the negative radial 
gradient has a physical origin, possibly related to the dilution of a metal-enriched gas outflow. 
Both interpretations are consistent with the apparent behavior of absolute and relative abundances. 


Although the IH/OH analysis revealed no clear trend for either temperature or metallicity, 
we proceeded with a finer annular sampling of the halo. This is desirable in order to obtain 
a radial profile of the gas density as a function of the emission measures (see later). Following 
the same adaptive criterion, we selected five regions within $r = 15$--80$\arcsec$ (H1--H5, 
with intermediate radii of 21.8, 30.8, 41.8, and 55.5$\arcsec$), containing $\sim$1800 net 
counts each at 0.5--1.5 keV (Table~\ref{t4}). Model C was applied 
in the spectral analysis, with temperatures free and abundances tied to their average values. 
We also made use of the deprojection technique available through the \textsc{xspec} model 
component \texttt{projct}. Indeed, since we can only probe the footprint of the emitting source 
on the plane of the sky, a system with a rough spherical symmetry is subject to some degree 
of radial mixing: the emission observed in a given annulus also includes a contribution from 
material located at larger distances. In this case, however, deprojection does not result in any 
significant difference, confirming the suggested lack of strong radial gradients. Only emission 
measures experience some variations, but the correction involved for the gas density is still 
limited ($<$~30\%), and therefore has been disregarded. 

\subsection{Azimuthal Analysis}
As discussed above, radial variations, if any, are limited. A possible, partial explanation 
is that the integration over the polar angle smooths out most of the existing differences. 
From Figs.~\ref{hd} and \ref{as}, in fact, one can easily appreciate that the surface brightness 
profile depends on the direction along which it is evaluated. Several structures, outstanding 
for intensity and/or extent, are confined in narrow azimuthal sectors, so that the relative 
information might be lost in a purely radial analysis. The aforementioned azimuthal regions 
S1--S6 (Fig.~\ref{hm}) were adaptively selected following the morphological appearance, and 
imposing that each key feature is sharply singled out, most notably the wings of the cross-like 
pattern. Each sector includes $\sim$1800 net counts at 0.4--2.5 keV, which represent 
$\sim$60--70\% of the total collected events (see Table~\ref{t4}). We then made use of 
two new models, F and G, both based on the simple \texttt{wabs*vapec} form. This choice 
is fully justified by our preliminary analysis, demonstrating that any additional spectral 
component is redundant, apart from a thermal one (which, however, cannot be explored 
over the individual zones).\footnote{The presence of localized absorption and non-thermal 
emission in the single regions has been ruled out by adding back into the model, one at a 
time, the corresponding spectral component.} 

\begin{table*}
\caption{Spatially-resolved Analysis over the Six Azimuthal Halo Subregions S1--S6.}
\label{t5}
\begin{scriptsize}
\begin{tabular}{cccccccccc}
\hline \hline
Reg & Mod & $kT$ & EM & $Z_\rmn{O}$ & $Z_\rmn{Ne}$ & 
$Z_\rmn{Mg}$ & $Z_\rmn{Si}$ & $Z_\rmn{Fe}$ & $C$-stat/$\nu$ \\
\hline 
S1 & F & 0.65$^{+0.10}_{-0.07}$ & 4.9$^{+2.2}_{-2.2}$ & 0.85$^{+1.54}_{-0.52}$ & 0.92$^{+1.35}_{-0.55}$ & 
0.98$^{+1.09}_{-0.46}$ & 0.40$^{+0.65}_{-0.39}$ & 0.16$^{+0.15}_{-0.06}$ & 119.5/103 \\
 & G & 0.66$^{+0.07}_{-0.05}$ & 6.5$^{+1.4}_{-1.3}$ & 0.60$^{+0.38}_{-0.28}$ & 0.52$^{+0.35}_{-0.28}$ & 
0.66$^{+0.29}_{-0.23}$ & 0.27$^{+0.32}_{-0.27}$ & 0.11$^{+0.04}_{-0.03}$ & 125.4/103\smallskip \\
S2 & F & 0.63$^{+0.11}_{-0.07}$ & 8.3$^{+2.1}_{-2.1}$ & 0.22$^{+0.40}_{-0.20}$ & 0.35$^{+0.36}_{-0.28}$ & 
0.20$^{+0.23}_{-0.17}$ & 0.48$^{+0.35}_{-0.25}$ & 0.10$^{+0.04}_{-0.03}$ & 114.4/114 \\
 & G & 0.67$^{+0.05}_{-0.05}$ & 6.5$^{+1.0}_{-0.9}$ & 0.42$^{+0.30}_{-0.23}$ & 0.51$^{+0.34}_{-0.27}$ & 
0.24$^{+0.20}_{-0.18}$ & 0.51$^{+0.31}_{-0.25}$ & 0.14$^{+0.04}_{-0.03}$ & 120.9/114\smallskip \\
S3 & F & 0.66$^{+0.07}_{-0.06}$ & 5.9$^{+2.0}_{-2.0}$ & 0.69$^{+0.85}_{-0.39}$ & 0.54$^{+0.65}_{-0.37}$ & 
0.56$^{+0.49}_{-0.28}$ & 0.54$^{+0.46}_{-0.30}$ & 0.14$^{+0.09}_{-0.04}$ & 83.3/108 \\
 & G & 0.66$^{+0.04}_{-0.05}$ & 6.5$^{+1.1}_{-1.0}$ & 0.64$^{+0.35}_{-0.27}$ & 0.37$^{+0.30}_{-0.25}$ & 
0.43$^{+0.23}_{-0.19}$ & 0.45$^{+0.28}_{-0.24}$ & 0.13$^{+0.04}_{-0.03}$ & 86.6/108\smallskip \\
S4 & F & 0.64$^{+0.09}_{-0.07}$ & 5.8$^{+2.4}_{-2.4}$ & 0.62$^{+1.01}_{-0.41}$ & 0.65$^{+0.93}_{-0.44}$ & 
0.58$^{+0.67}_{-0.33}$ & 0.31$^{+0.55}_{-0.31}$ & 0.16$^{+0.13}_{-0.06}$ & 102.8/99 \\
 & G & 0.64$^{+0.05}_{-0.05}$ & 6.6$^{+1.2}_{-1.1}$ & 0.60$^{+0.39}_{-0.28}$ & 0.41$^{+0.34}_{-0.27}$ & 
0.37$^{+0.24}_{-0.21}$ & 0.18$^{+0.32}_{-0.18}$ & 0.15$^{+0.04}_{-0.04}$ & 108.7/99\smallskip \\
S5 & F & 0.82$^{+0.10}_{-0.07}$ & 6.1$^{+1.7}_{-1.7}$ & 0.78$^{+1.03}_{-0.47}$ & 0.23$^{+0.66}_{-0.23}$ & 
0.47$^{+0.43}_{-0.29}$ & 0.19$^{+0.27}_{-0.19}$ & 0.14$^{+0.10}_{-0.05}$ & 102.8/117 \\
 & G & 0.78$^{+0.05}_{-0.05}$ & 7.0$^{+1.7}_{-1.5}$ & 0.34$^{+0.28}_{-0.21}$ & 0.88$^{+0.42}_{-0.32}$ & 
0.57$^{+0.25}_{-0.21}$ & 0.34$^{+0.26}_{-0.24}$ & 0.09$^{+0.03}_{-0.02}$ & 121.0/117\smallskip \\
S6 & F & 0.65$^{+0.14}_{-0.07}$ & 7.4$^{+2.3}_{-2.6}$ & 0.55$^{+0.96}_{-0.32}$ & 0.67$^{+0.70}_{-0.35}$ & 
0.21$^{+0.30}_{-0.21}$ & 0.74$^{+0.51}_{-0.36}$ & 0.07$^{+0.04}_{-0.03}$ & 106.5/114 \\
 & G & 0.75$^{+0.05}_{-0.05}$ & 5.5$^{+1.2}_{-0.9}$ & 0.64$^{+0.39}_{-0.28}$ & 0.86$^{+0.43}_{-0.33}$ & 
0.37$^{+0.23}_{-0.20}$ & 0.88$^{+0.38}_{-0.32}$ & 0.08$^{+0.02}_{-0.03}$ & 121.9/114 \\
\hline
\end{tabular}
\end{scriptsize}
\flushleft
\small{\textit{Note.} The basic spectral form is \texttt{wabs*vapec} for both model F and G, 
but the fitting procedure is different (see the text for details). All the physical quantities and 
their respective units have been defined in Table~\ref{t3}.}
\end{table*}

\begin{figure}
\includegraphics[width=15cm]{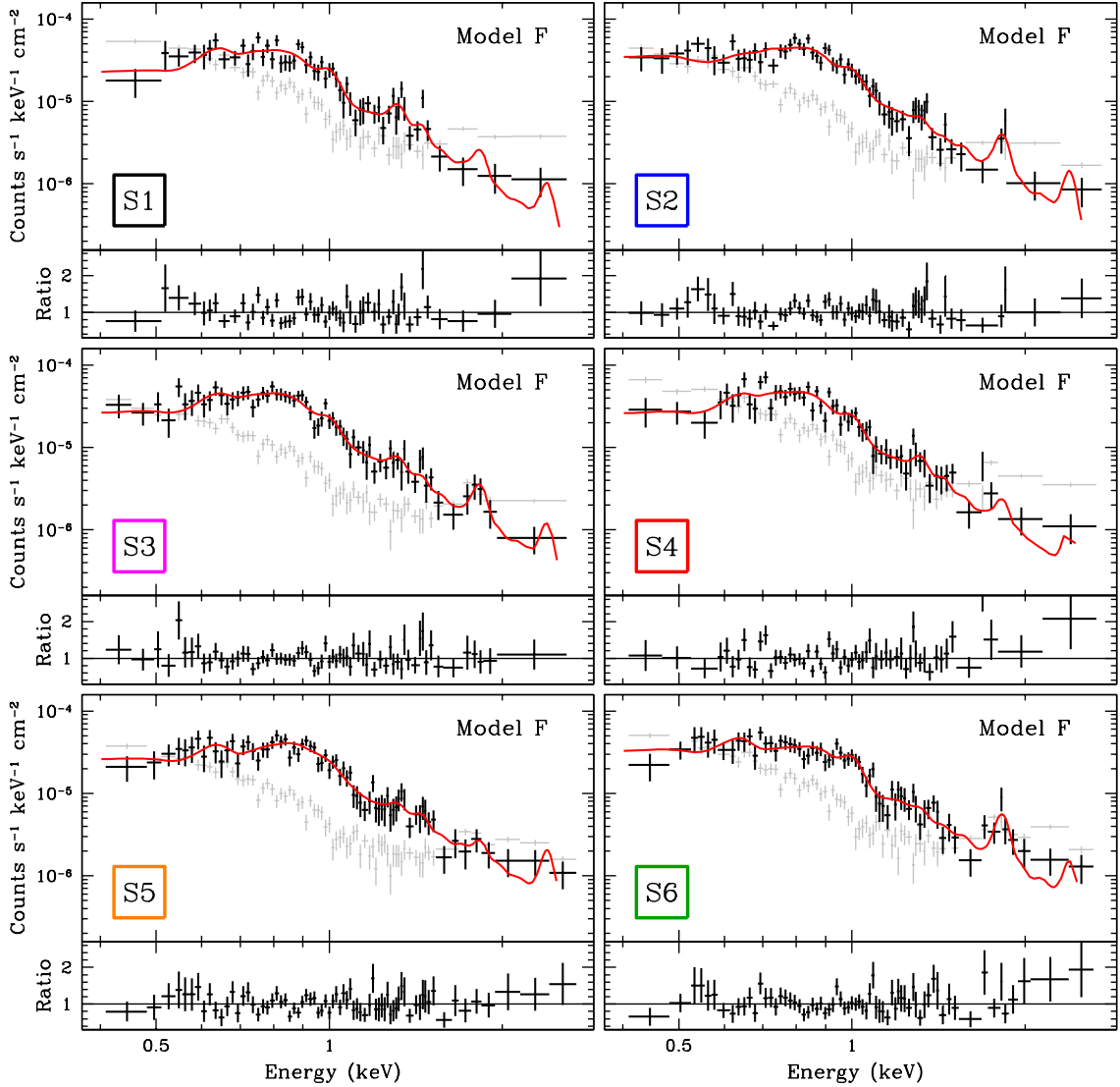}
\caption{Source spectra, best-fitting models and data/model ratios for the six azimuthal 
regions S1--S6. The background is also outlined (in light grey) for comparison. The same 
graphical rebinning (to at least a 2.5$\sigma$ significance per channel) has been applied to all the spectra, while the model curves retain the intrinsic 
energy resolution. This elucidates the necessity of dealing with ungrouped spectra, adopting 
$C$-stat in each halo subregion, in order to constrain with higher accuracy the elemental 
abundances.} 
\label{fm}
\end{figure}

\begin{figure}
\includegraphics[width=15cm]{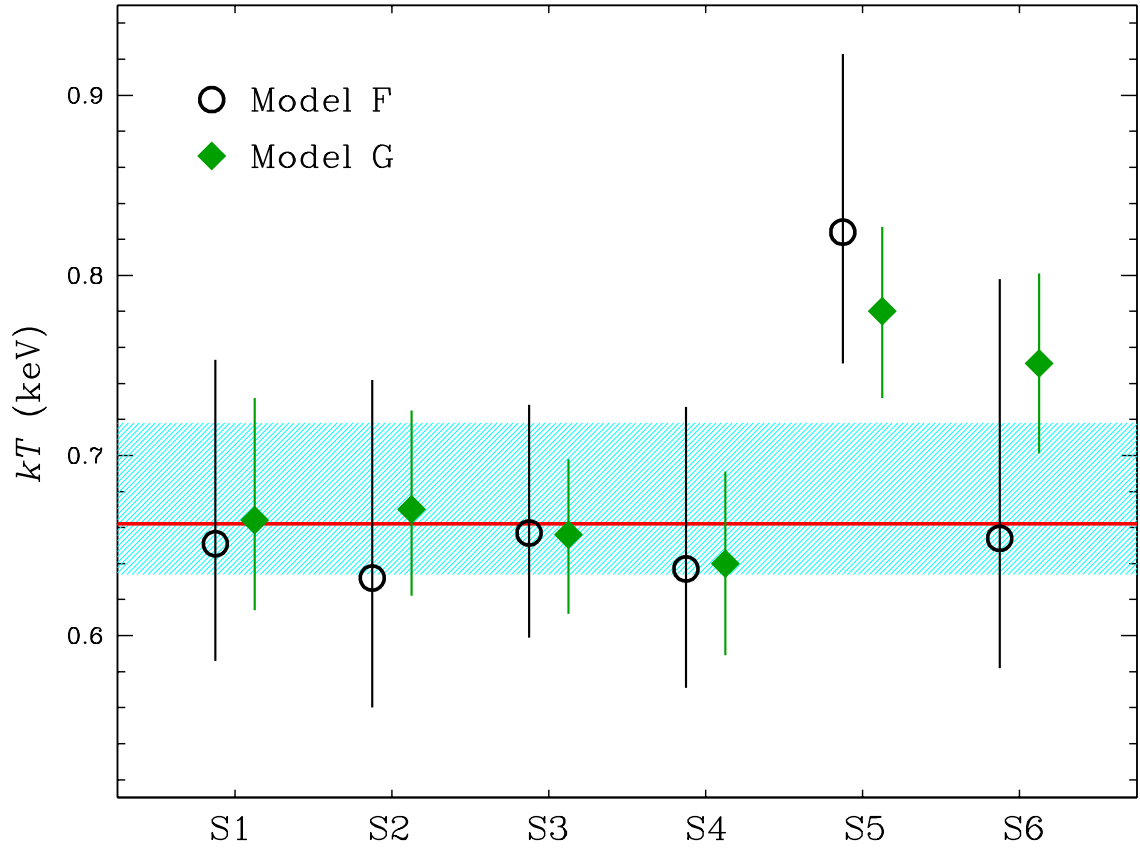}
\caption{Comparison of the gas temperatures in the S1--S6 subregions obtained from models 
F (black circles) and G (green diamonds). The best-fitting spectroscopic temperature (solid line) 
and its confidence range (shaded area) from model C, averaged over the full halo, are also shown 
for reference.}
\label{kt}
\end{figure}

\begin{figure}
\includegraphics[width=15cm]{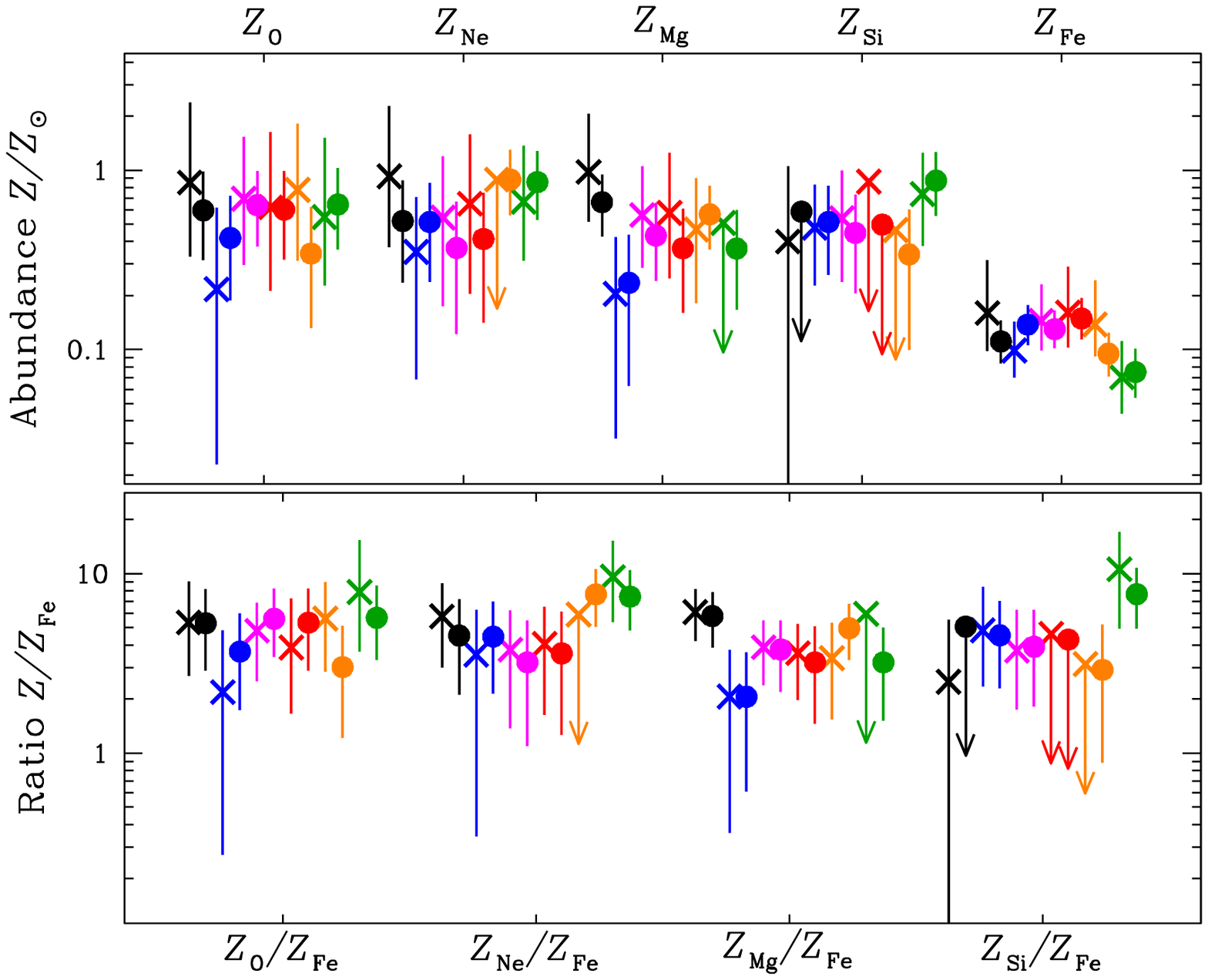}
\caption{Best-fitting elemental abundances (top) and abundance ratios (bottom) in the six azimuthal 
sectors of the halo. The color code is the same defined in Fig.~\ref{hm}. Results from both model F 
(crosses) and model G (dots) are plotted (see Tables~\ref{t5} and \ref{t7}).}
\label{ar}
\end{figure}

At first, we took the most general approach of dealing with each subregion independently 
(model F). The six spectra and their best fits are shown in Fig.~\ref{fm}. The results for the 
individual gas temperatures and abundances are summarized in Table~\ref{t5}, and plotted 
in Figs.~\ref{kt} and \ref{ar}. Evidence for a slightly hotter gas component is found in S5, 
and also abundance measures provisionally hint at sizable gradients. The error bars are quite 
large, though. With model G, we have therefore adopted an alternative strategy to reduce both 
the systematic and the statistical uncertainties on absolute and relative abundances. The basic 
idea is to take advantage of the fact that the extent of any variation across the halo has proven 
to be rather limited. This allowed us to extract some additional information by investigating 
the behavior of each of the key spectral parameters separately, with all the other variables 
tied (but not frozen) to their average value. 


We created in turn six spatial maps for $kT$, and O, Mg, Ne, Si, Fe abundances by performing 
a simultaneous fit of the S1--S6 regions, with the single observable of interest free to vary in 
the different spectra and the other parameters (apart from the normalization) tied to a common 
value. According to this method, Table~\ref{t5} should be read along rows for model F only. 
The physical information from model G is conveyed by columns, which correspond to a specific 
temperature or abundance map. Since no quantity is frozen, every fit delivers an independent 
set of average values for the collateral parameters. Not surprisingly, these are always coincident 
(typically within $\sim$5--10\%) with the results of model C, where temperatures and abundances 
are all tied over S1--S6, and from which the entries for emission measure and $C$-stat (evaluated 
on the single region of interest) have been taken.

\begin{figure}
\includegraphics[width=15cm]{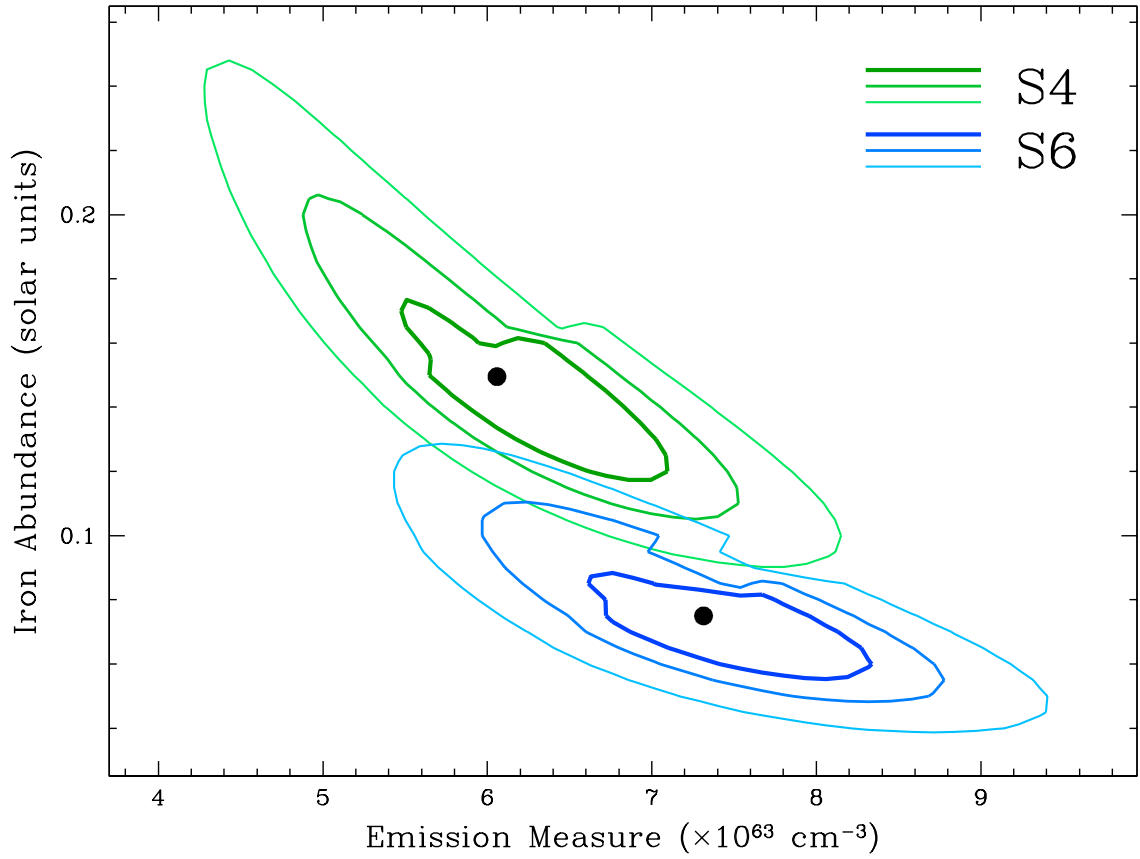}
\caption{$Z_\rmn{Fe}$--EM contour plots for S4 and S6, the two regions in the halo with the 
larger gradient of iron abundance. The different curves correspond to the nominal 68, 90 and 
99\% confidence levels for model G.} 
\label{ic}
\end{figure}

Model G effectively returns smaller error bars, as illustrated in Figs.~\ref{kt} and \ref{ar}. This 
result is not actually achieved at the expense of accuracy. The measure of individual abundances, 
in first approximation, relies on a specific, narrow energy range, whose continuum is now 
constrained over the full halo. The fact that $C$-stat undergoes just tiny increments with 
respect to model F, when assessed over the single regions, implies that the entire soft X-ray 
spectrum is well reproduced, not only the band of interest for each abundance map. This 
further confirms the reliability of our approach. The last sensitive issue to investigate is the 
possible degeneracy between iron abundance and emission measure, which can partly weaken 
the significance of any gradient. In spite of the low $Z_\rmn{Fe}$ value, iron L-shell 
contribution is still dominant, and it is difficult to disentangle from the continuum at CCD 
resolution. In Fig.~\ref{ic} we show the $Z_\rmn{Fe}$--EM confidence contours for the two 
regions (S4 and S6) with the larger difference (2.8$\sigma$) in $Z_\rmn{Fe}$, proving that 
such effect is minimal. A gradient of iron abundance in the halo is then detected with a 
$\sim$2$\sigma$ significance. With little dependence on the specific model (see Fig.~\ref{ar}), 
a similar result possibly holds for magnesium, while for all the other $\alpha$-elements error 
bars are still too large to draw any conclusion. 

\subsection{Physical Parameters}
The key physical properties (density, mass, pressure, thermal energy) of the X-ray emitting 
gas can be inferred from the results of the spectral analysis, provided that the volume of the 
halo is estimated. We assumed that the thickness of the system is comparable to its width. 
This is a reasonable approximation for a merger about to enter the final coalescence stage 
like NGC~6240. Moreover, the main galactic body in the optical is reminiscent of a distorted 
edge-on disk (Gerssen et al. 2004), implying a substantial span along the line of sight. 
A nearly face-on configuration (like in the case of the Antennae, NGC~4038/4039; Fabbiano 
et al. 2003) does not seem to apply here, and can be safely ruled out. Based on this conjecture, 
we computed the volume $\mathcal{V}$ of the halo as follows:
\[
\mathcal{V} = \sum \frac{2 \theta_i}{3} R_i^3 \left( \frac{2 \mathcal{S}_i}{\theta_i R_i^2} \right)^{3/2} 
\simeq 1.9 \times \sum \mathcal{S}_i^{3/2} \theta_i^{-1/2},
\]
where $\mathcal{S}_i$ is the surface derived for each of the six azimuthal regions S1--S6 
from the 3$\sigma$ contour of Fig.~\ref{hd}, and $R_i$ and $\theta_i$ are the outer radius 
(in kpc) and the central angle (in radians) of the individual sectors. We eventually obtained 
that $\mathcal{V} \simeq 2.15 \times 10^5$ kpc$^3$. In the detailed calculations we have 
also taken into account the geometrical projection effects. This is mainly relevant to the 
volumes of the annular regions H1--H5, which strongly depend on the radius of the 
embedding sphere, taken to be 100$\arcsec$ ($\pm 20\arcsec$). Such correction is 
negligible (roughly 1\%) on $\mathcal{V}$, also in the light of the large systematic 
uncertainties.

\begin{table}
\caption{Global and Local Physical Properties of the Hot Gas.} 
\label{t6}
\begin{tabular}{cccccccc}
\hline\hline
Reg & Mod & $\mathcal{V}$ & $n_e$ & $M_g$ & $p$ & $E_\mathit{th}$ & $\tau_\rmn{c}$ \\
\hline
FH & B & 6.31 & 2.5$^{+0.2}_{-0.3}$ & 13.1$^{+1.2}_{-1.2}$ & 5.2$^{+0.7}_{-0.6}$ & 
49$^{+7}_{-5}$ & 3.7$^{+1.0}_{-0.6}$ \\
H1 & C & 0.54 & 3.6$^{+0.5}_{-0.4}$ & 1.7$^{+0.2}_{-0.2}$ & 8.3$^{+1.1}_{-1.1}$ & 
6.7$^{+0.9}_{-0.9}$ & 2.6$^{+0.3}_{-0.4}$ \\
H2 & C & 1.00 & 2.6$^{+0.4}_{-0.3}$ & 2.2$^{+0.3}_{-0.3}$ & 5.6$^{+0.9}_{-0.7}$ & 
8.5$^{+1.3}_{-1.2}$ & 3.3$^{+0.6}_{-0.4}$ \\
H3 & C & 1.63 & 2.1$^{+0.3}_{-0.3}$ & 2.8$^{+0.4}_{-0.3}$ & 4.4$^{+0.7}_{-0.7}$ & 
10.7$^{+1.6}_{-1.6}$ & 4.2$^{+0.7}_{-0.6}$ \\
H4 & C & 2.48 & 1.7$^{+0.2}_{-0.3}$ & 3.5$^{+0.5}_{-0.5}$ & 3.7$^{+0.6}_{-0.6}$ & 
13.7$^{+2.2}_{-2.2}$ & 5.5$^{+0.9}_{-1.0}$ \\
H5 & C & 2.95 & 1.5$^{+0.4}_{-0.3}$ & 3.8$^{+0.7}_{-0.9}$ & 3.0$^{+0.8}_{-0.7}$ & 
13.2$^{+2.9}_{-3.5}$ & 5.3$^{+1.2}_{-1.4}$ \\
S1 & F & 1.02 & 2.2$^{+0.4}_{-0.6}$ & 1.9$^{+0.4}_{-0.5}$ & 4.5$^{+1.2}_{-1.2}$ & 
7.0$^{+1.8}_{-1.9}$ & 3.4$^{+0.9}_{-1.0}$ \\
S2 & F & 1.06 & 2.8$^{+0.3}_{-0.4}$ & 2.5$^{+0.3}_{-0.3}$ & 5.7$^{+1.2}_{-1.0}$ & 
9.0$^{+1.9}_{-1.6}$ & 3.9$^{+0.8}_{-0.7}$ \\
S3 & F & 0.67 & 3.0$^{+0.4}_{-0.6}$ & 1.7$^{+0.2}_{-0.3}$ & 6.3$^{+1.2}_{-1.3}$ & 
6.3$^{+1.2}_{-1.3}$ & 2.8$^{+0.6}_{-0.5}$ \\
S4 & F & 1.25 & 2.1$^{+0.4}_{-0.5}$ & 2.3$^{+0.4}_{-0.6}$ & 4.4$^{+1.0}_{-1.1}$ & 
8.2$^{+2.0}_{-2.1}$ & 3.8$^{+1.0}_{-1.0}$ \\
S5 & F & 0.99 & 2.5$^{+0.3}_{-0.4}$ & 2.1$^{+0.3}_{-0.3}$ & 6.5$^{+1.2}_{-1.1}$ & 
9.7$^{+1.8}_{-1.6}$ & 4.4$^{+0.8}_{-0.8}$ \\
S6 & F & 1.32 & 2.4$^{+0.3}_{-0.5}$ & 2.6$^{+0.4}_{-0.5}$ & 5.0$^{+1.3}_{-1.1}$ & 
9.9$^{+2.5}_{-2.2}$ & 4.3$^{+1.1}_{-1.0}$ \\
\hline
\end{tabular}
\flushleft
\small{\textit{Note.} $\mathcal{V}$: estimated volume in 10$^{69}$ cm$^3$; $n_e$: electron density 
in 10$^{-3}$ cm$^{-3}$; $M_g$: total mass in 10$^9 M_{\sun}$; $p$: pressure in 10$^{-12}$ dyne 
cm$^{-2}$; $E_\mathit{th}$: thermal energy in 10$^{57}$ erg; $\tau_\rmn{c}$: cooling time in Gyr.}
\end{table}

It is actually the effective volume occupied by the gas that matters for the estimate of 
the physical properties. This is related to the geometrical volume through the expression 
$V_\rmn{eff} = \eta \mathcal{V}$, where $\eta$ encompasses both the filling factor of 
the emitting gas and any correction scale factor. Assuming $n_e \sim n_\rmn{H}$, the 
emission measure can be written as $\rmn{EM} \simeq n_e^2 \eta \mathcal{V}$, from 
which we extracted the electron density, and hence the gas pressure $p=2 n_e kT$, mass 
$M = n_e \eta \mathcal{V} m_p$ and thermal energy $E_\mathit{th} = 3 n_e \eta \mathcal{V} kT$. 
The cooling time $\tau_\rmn{c}$ was estimated as the ratio between the thermal energy 
content and the 0.3--8 keV thermal luminosity, inferred from the spectral fits. Our results 
for the full halo and its different subregions are presented in Table~\ref{t6}, and their main 
implications are discussed in the next Section. 

All the physical quantities mentioned above contain the unknown filling factor $\eta$, which 
cannot be directly constrained. Hydrodynamic simulations suggest that the low-density, hot 
gas component ($n_e \sim 10^{-2}$ cm$^{-2}$, $kT \sim 10^7$ K) in starburst-driven winds 
is volume-filling ($\eta \sim 0.7$), even if its contribution to the soft X-ray emission may not 
be dominant (e.g., Strickland \& Stevens 2000). Since the dependence from $\eta$ is mild 
($n_e$, $p \propto \eta^{-1/2}$, while $M_g$, $E_\mathit{th}$, $\tau_\rmn{c} \propto \eta^{1/2}$), 
the estimates given in Table~\ref{t6} correspond to $\eta = 1$, keeping in mind that the actual 
values can differ by a factor of two or more. 

\section{Discussion}
The smoothed soft X-ray images of NGC~6240 reveal a spectacular halo of 
hot gas, with an average radial extent of $\sim$50 kpc. Its total luminosity in 
excess of 4$\times$10$^{41}$ erg s$^{-1}$ is comparable to that of small groups 
of galaxies (Mulchaey 2000) and giant ellipticals (Canizares et al. 1987; Mathews 
\& Brighenti 2003), while its age can be provisionally estimated through the 
sound-crossing time. The adiabatic sound speed in a gas with $kT \simeq 0.65$ 
keV (7.5 million K), and mean molecular weight $\mu \simeq 0.6$ (as per a 
fully-ionized, primordial plasma), is $c_s = (5kT/3\mu m_p)^{1/2} \simeq 420$ 
km s$^{-1}$, resulting in a dynamical age of $\sim$200 Myr. In this Section we 
consider thoroughly all the viable physical explanations for the observed halo 
luminosity, size, temperature, morphology, and metallicity, in order to achieve 
a self-consistent interpretation, and discuss the origin and possible evolution 
of the system. 

\subsection{Galaxy Merger}
Close interactions and mergers represent the most critical stage of 
galaxy evolution over cosmic time. Both models and observations 
indicate that, after the collision of a pair of massive, gas-rich spirals, 
the kinematic and structural properties of the remnant are consistent 
with elliptical galaxies (e.g., Barnes 1988; Dasyra et al. 2006). The 
relation between mergers and hot gaseous halos, instead, is still 
rather controversial. The first encounter of the progenitor disks 
may lead to substantial heating, through the dissipation of the kinetic 
energy in the gas components and the formation of copious shocks in 
the contact layers. In the present case, the former mechanism is largely 
insufficient. Assuming two identical progenitors, the amount of kinetic 
energy deposited during the merger is roughly $M_g v_c^2/8$, where 
$M_g$ is the mass of the X-ray emitting gas and $v_c$ is the relative 
speed during the collision. The thermal energy content of the halo 
($E_\mathit{th} \simeq 5 \times$10$^{58}$ erg) corresponds to $v_c 
\sim 1200$ km s$^{-1}$, which is nearly twice the velocity observed 
in the Taffy Galaxies (Braine et al. 2003), a characteristic example of 
a head-on collision.\footnote{Note that this estimate is independent 
of the filling factor and the actual values of $E_\mathit{th}$ and $M_g$, 
since both quantities vary as $\eta^{1/2}$.}

On the other hand, the same halo luminosity is probably too large for 
the interaction of two galaxies with properties similar to the Milky Way. 
The peak of the X-ray luminosity in galaxy mergers is attained with the 
final coalescence, or shortly before (Brassington et al. 2007). According 
to simulations, its value depends on several factors, among which the 
size and the gas fraction of the progenitors, the orbital parameters, and 
the relative orientation. The overall complexity is a possible reason for 
which extended X-ray emission is not systematically found in mergers.
In principle, also the presence of a supermassive black hole has an 
important role in the gas heating and its subsequent dispersion (Cox et 
al. 2006).  In NGC~6240, the AGN pair is believed to contribute significantly 
to the bolometric luminosity (Lutz et al. 2003; Egami et al. 2006), yet their 
direct feedback on the circumnuclear environment is difficult to establish. Their 
impact has presumably been negligible so far, as both nuclei are still completely 
enshrouded by a thick shell of dust and gas (Iwasawa \& Comastri 1998; Vignati 
et al. 1999). In this view, unless the aforementioned physical parameters are 
unusually fine-tuned, a maximum $L_\rmn{X} > 10^{41}$ erg s$^{-1}$ can 
be reached only if the progenitors are much more massive than the Milky 
Way (see also below). This makes of NGC~6240 an exceptional case study with 
respect to both the observations of known systems and the numerical simulations 
of mergers. Accepting that NGC~6240 is now at its peak X-ray luminosity, in fact, 
it should be kept in mind that the entire, diffuse X-ray emission is $\sim$3 times 
that of the halo alone.

The tidal forces can be responsible for the presence of hot gas at large 
distance from the center. The arms of the wide, cross-like morphological 
feature are openly suggestive of the typical tidal tails observed in mergers, 
but they have no detected counterpart at different wavelengths. 
Another problematic point resides in the gas temperature. When 
swept by an adiabatic shock, the gas is heated to $kT_\rmn{s} = 3 
\mu m_p v_\rmn{s}^2/16$, where $v_\rmn{s}$ is the speed of the 
shock front (Hollenbach \& McKee 1979). The velocity required to 
produce a temperature of 0.65 keV is $v_\rmn{s} \sim 750$ km s$^{-1}$, 
which is still too high when compared to the inferred orbital velocity of the 
two nuclei of NGC~6240 (155 km s$^{-1}$; Tecza et al. 2000), even if the 
rotation of the parent disks is taken into account. Finally, the inelastic 
nature of the galaxy merger does not seem capable of accounting for 
the energetics of the X-ray halo. Yet, the ongoing nuclear starburst is 
itself triggered by the global redistribution of the gas components that 
follows the strong gravitational disturbances, and represents a much 
more efficient energy source. 

\subsection{Starburst Wind}
During a starburst, the formation of young, massive stars persists 
until the depletion of the molecular gas reservoir and/or the onset 
of some self-regulation mechanism. This activity leads to intense 
mass loss, stellar winds and frequent supernova (SN) explosions. 
The resultant deposition of energy, momentum and metals into 
the intergalactic medium is a key driver of galaxy evolution. Based 
on the stellar $K$-band emission of NGC~6240, Tecza et al. (2000) 
evinced that a short ($\sim$5 Myr) burst of star formation took place 
$\sim$20 Myr ago, possibly coincident with the latest perigalactic 
passage (but see Engel et al. 2010b). As the mechanical energy injected 
by a starburst is roughly 1\% of its bolometric luminosity (e.g., Leitherer 
et al. 1999), and the total luminosity of NGC~6240 is $L_\rmn{bol} 
\approx L_\rmn{IR} \sim 3 \times$10$^{45}$ erg s$^{-1}$ (Armus 
et al. 2006), it takes at least $\sim$50 Myr for the starburst to supply 
the estimated thermal energy content of the X-ray halo, neglecting 
the AGN contribution to $L_\rmn{IR}$. It is therefore highly questionable 
whether the halo is a direct consequence of the most recent starburst 
episode, also considering its huge size, which calls for an average outflow 
velocity of $\sim$2500 km s$^{-1}$ over the last 20 Myr. This would be 
actually consistent with the fastest shocks we identified in the central 5 
kpc (Wang et al. 2013a), so the key issue is whether an outflow could 
sustain a similar expansion rate for several tens of kpc. 

In the classical model of starburst-driven winds (Chevalier \& Clegg 1985), 
the terminal outward velocity is $v_\infty = (2 \dot{E}/\dot{M})^{1/2}$, 
where $\dot{E}$ and $\dot{M}$ are the energy and mass injection rates, 
respectively. By adopting a standard mechanical energy input of 10$^{51}$ 
erg per SN (Chevalier 1977), and an average SN mass of $\sim$10 $M_{\sun}$ 
with intrinsic mass deposition fraction of $\sim$10\%, we can write 
$v_\infty \sim 10^4 (\xi/\Lambda)^{1/2}$ km s$^{-1}$, where the 
thermalization efficiency $\xi$ accounts for any radiative losses, and the 
mass-loading factor $\Lambda$ is the ratio between the total mass of the 
gas heated within the starburst and the mass of the direct SN ejecta 
(Veilleux et al. 2005). The obvious effect of radiative losses and mass loading 
is to slow down the wind. In the absence of observational constraints for 
$\xi$ and $\Lambda$, the previous expressions can be rearranged further 
into the fully equivalent $v_\infty = (5kT_c/\mu m_p)^{1/2}$, which brings 
out the dependence on the central gas temperature only (Strickland \& 
Heckman 2009). By approximating $T_c$ with the nearly constant halo 
temperature, we obtain that $v_\infty \simeq 720$ km s$^{-1}$, 
commensurate with the average width of the optical line-emitting gas 
(Heckman et al. 1990), and with the terminal velocity of the nuclear 
outflow determined from the blueshifted Na~\textsc{i} $\lambda 
\lambda$5890, 5896 doublet (Na D) absorption (Heckman et al. 2000). 

Alternatively, we assume that $\xi \simeq 1$, and attempt to assess 
$\Lambda$ for a constant mass injection rate since the putative epoch 
of the most recent starburst. The SN rate in NGC~6240, as derived from 
the non-thermal radio continuum emission, is $\sim$2 yr$^{-1}$ (van 
der Werf et al. 1993; Beswick et al. 2001). This value strongly depends 
on the star formation history adopted (continuous, instantaneous or 
merger-induced), and is possibly overestimated by up to an order of 
magnitude (Engel et al. 2010b). At least, it represents a useful upper limit 
for the present purpose.\footnote{Interestingly, among the various 
estimates considered in Wang et al. (2013a), we obtained a SN rate of 
$\sim$3 yr$^{-1}$, using the correlation between 
[Fe~\textsc{ii}] at 1.26 $\mu$m and SN rate found in nearby starburst 
galaxies (Rosenberg et al. 2012).} A total mass of hot gas in the halo of 
$\sim$10$^{10} M_{\sun}$ (Table~\ref{t6}) indicates that $\Lambda \sim 
300$. Taken at the face value, this would imply a mass loading efficiency 
(i.e., the mass outflow rate normalized to the star formation rate) larger 
than one, which is not unusual among ULIRGs (Rupke et al. 2005), and 
is comparable to that inferred for similar IR-luminous, merging systems 
like Mrk~266 (Wang et al. 1997) and Arp~299 (Heckman et al. 1999). 
In both of the latter cases, however, the dynamical age of the X-ray 
nebula is broadly consistent with the typical starburst lifetime. In 
NGC~6240, such a substantial mass loading would correspond to 
$v_\infty \simeq 700$ km s$^{-1}$, in excellent agreement with our 
previous estimate. In conclusion, any galactic wind originating some 20 
Myr ago must have traveled for $\sim$10--15 kpc at most. Remarkably, 
this is just beyond the size of the optical nebula and of the soft X-ray core.

The two equivalent expressions for the terminal velocity used above are 
derived from wind models that neglect both ambient gas and gravitational 
forces. The mass-loading is centralized (i.e., it takes place within the starburst 
region itself), and is typically limited to a few $M_{\sun}$ per SN ($\Lambda < 10$; 
Suchkov et al. 1996). The archetypal source to which these analytical models and 
hydrodynamical simulations are compared is M82, since it is isolated, contains no 
AGN, and the hot gas is free to escape perpendicular to the galactic plane (Griffiths 
et al. 2000). The huge mass loading required for mergers, instead, implies the 
entrainment of tens to hundreds of $M_{\sun}$ per SN in the ambient material 
swept by the wind. In a ULIRG-like system, due to the high density and the disturbed 
geometry, it is plausible that the conversion of mechanical energy into radiation 
is much less efficient than in M82-like starbursts (e.g., Colina et al. 2004), and 
even that outflows expand only over little distances before terminating. However, 
as the clumpy morphological structures in NGC~6240 reach out to the largest 
physical scales, a wind nature for the entire halo cannot be conclusively rejected. 
Enhanced star formation with respect to quiescent galaxies has been likely in 
place since the beginning of the interaction, dated to several hundreds of Myr 
ago. Moreover, in these earlier times the additional momentum injection due 
to the AGN radiation pressure could have been non negligible, with a possible 
boost effect on the starburst outflows.
 
\begin{figure}
\includegraphics[width=15cm]{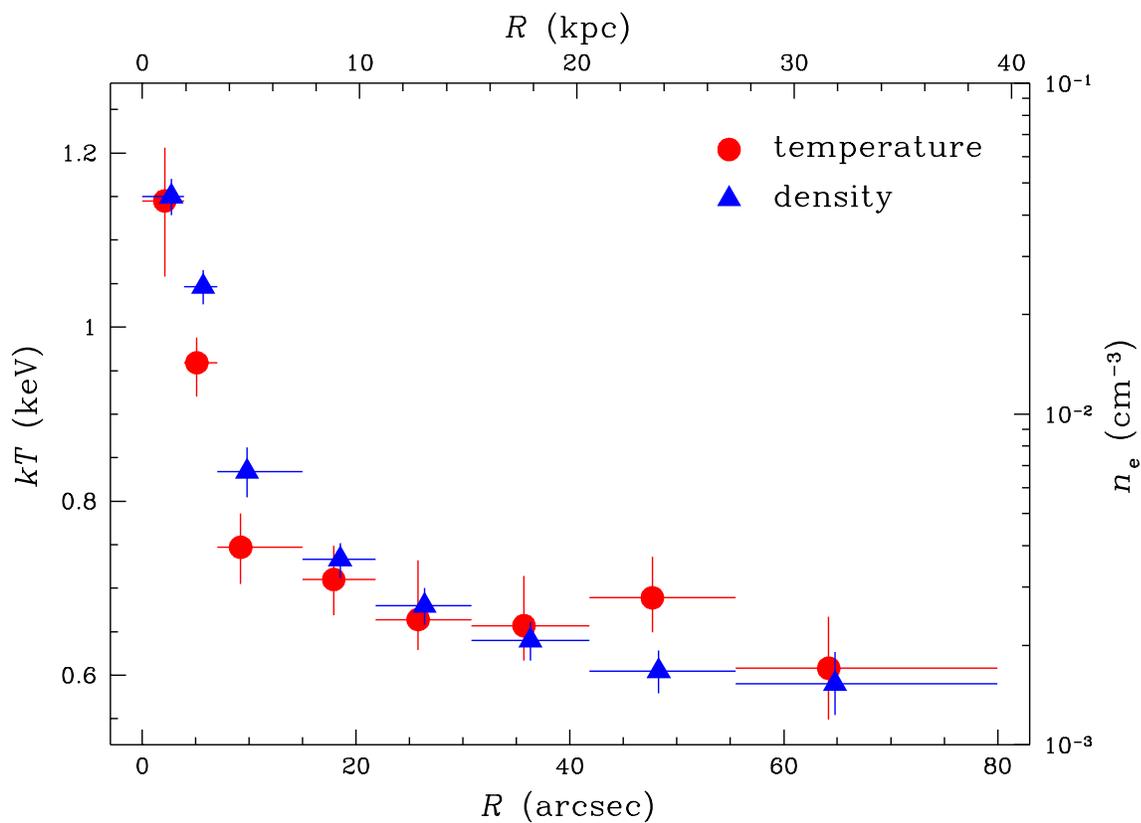}
\caption{Radial dependence of temperature (red dots) and electron density (blue crosses) of 
the soft X-ray emitting gas in NGC~6240, estimated on different annular regions up to a 
galactocentric distance of 80$\arcsec$ ($\sim$40 kpc). Note the logarithmic scale adopted 
for densities on the right-hand vertical axis.}
\label{td}
\end{figure}

The \textit{secular} wind scenario, in keeping with the dynamical age of 
$\sim$200 Myr, can be probed through the average halo properties. In particular, 
the gas density in a freely expanding wind should decrease as $\sim r^{-2}$, as a 
consequence of the mass conservation law with constant outflow velocity at large 
distances. If the gas is nearly isothermal, the observed luminosity is related to the 
emission measure in a straightforward manner, hence the surface brightness is 
expected to exhibit a power-law dependence on radius as well: qualitatively, 
$\Sigma \propto n_e^2 \mathrm{d}V/\mathrm{d}S \sim r^{-3}$. Incidentally, 
this behavior was first observed in M82 with \textit{Einstein} (Fabbiano 1988). 
We have then fitted the radial profile of Fig.~\ref{sb} at distances $r > 12$ kpc, 
to avoid any contamination from the core component. As already indicated by 
the large scale radius in the $\beta$-model fit (Table~\ref{t2}), a power-law 
shape gives a very poor description ($\chi^2/\nu \sim 171/24$) of the average 
surface brightness. Indeed, over the range explored, $\Sigma(r)$ follows very 
well ($\chi^2/\nu \simeq 16/24$) an exponential decline with scale length 
$l = 10.1(\pm 0.2)$ kpc. The temperature and density profiles are plotted in 
Fig.~\ref{td}. Although only five measurements are available from the spectral 
fits to regions H1--H5, and the relative uncertainties are quite large, the halo 
gas density decreases with radius as $n_e \propto r^{-\alpha}$ with 
$\alpha \simeq 0.75(\pm 0.15)$, showing a milder trend with respect to a 
single adiabatic flow. The temperature is consistent with being constant 
beyond $\sim$5 kpc. These features could be explained by the superposition 
of successive winds emanating from different locations, suggesting a 
widespread rather than centrally-concentrated starburst.

For completeness, we have extended the $kT$ and $n_e$ profiles to smaller 
radii by extracting the spectra from three annular regions in the core, which 
provides twice as many counts as the halo over the 0.5--1.5 keV energy range. 
In this case, our single thermal model is purely phenomenological, since a 
multi-temperature plasma is known to be present in the nuclear environment 
(Netzer et al. 2005; Wang et al. 2013a). This was partly compensated through 
the power-law component, whose photon index was left free to vary and 
constrained up to 5 keV. Moderate column densities (a few $\times$10$^{21}$ 
cm$^{-2}$ around the nuclei, then reducing with distance) were also required. 
In brief, within the inner regions the gas density becomes compatible with the 
$r^{-2}$ trend, while a central temperature $T_c \sim 1$--1.5 keV does not 
substantially modify the wind expansion range suggested above. This simplified, 
preliminary analysis confirms that the soft X-ray emission in the core of NGC~6240, 
which is spatially correlated with the H$\alpha$ filaments, is most likely dominated 
by shock-heated material in a recent starburst-driven wind, whose connection 
with the simultaneous black hole growth (e.g., King 2005) is hard to determine. 
As the detailed study of this component is the subject of a companion paper (Wang 
et al. 2013b), its physical properties are not discussed further in this context.

\subsection{Dark Matter}
The large deviations from a classical wind model, in particular the flat 
temperature distribution, hint at near-isothermal, hydrostatic equilibrium. 
Together with the long cooling time (a few Gyr), this implies some form 
of gravitational confinement. Given the extent of the hot-gas halo, the 
baryonic matter is likely just a fraction of the total gravitating mass. 
Assuming the existence of a dark matter halo with an NFW density profile 
(Navarro et al. 1997), the gas virial temperature can be written as follows: 
\[
kT_\rmn{vir} = \frac{1}{2} \mu m_p \frac{GM_\rmn{vir}}{r_\rmn{vir}} 
\simeq 0.135 \left(\frac{M_\rmn{vir}}{10^{12}M_{\sun}}\right) \left(\frac{r_\rmn{vir}}{100~\rmn{kpc}}\right)^{-1} ~\rmn{keV},
\]
where $r_\rmn{vir}$ is the virial radius, usually taken to enclose an 
overdensity by a factor of 200 with respect to either the mean or the 
critical cosmic value at the time of the gravitational collapse, and 
$M_\rmn{vir}$ the virial mass within the corresponding volume, serving 
as a proxy for the total halo mass. Accordingly, the dark halo of NGC~6240 
has to be both massive and compact in order to confine the soft X-ray emitting 
gas within its potential well. Making use of the relations in Mo \& White (2002) 
we can render explicit the dependence of $r_\rmn{vir}$ on mass and redshift, 
obtaining that $kT_\rmn{vir} \sim 0.042~M_{12}^{2/3} (1+z)$ keV, where 
$M_{12}$ is the virial mass expressed in units of 10$^{12} M_{\sun}$. 

A possible evolutionary scenario consistent with the observed gas temperature 
of 0.65 keV is that of a cold dark matter halo formed at $z \sim 2$, with a total 
mass of $\sim$10$^{13} M_{\sun}$. This value marks the standard separation 
between galaxy-scale and group-scale halos (see Humphrey et al. 2006). For 
comparison, the Milky Way is estimated to have a virial mass of 
$\sim$1.3$\times$10$^{12} M_{\sun}$ (McMillan 2011). NGC~6240 is then 
consistent with the central remnant of a group of galaxies. This interpretation 
was first proposed by Huo et al. (2004), considering the huge content of hot 
gas ($> 10^{10} M_{\sun}$), the low metallicity of the outer halo, and the large 
velocity dispersion (the full width at zero intensity of CO emission lines reaches 
up to $\sim$1400 km s$^{-1}$; Feruglio et al. 2012). However, no evidence has 
been reported so far of a present-day galaxy overdensity in the surroundings of 
NGC~6240. Any other group member may have experienced a merger in the past, 
but in this case the progenitors of the final coalescence we are now witnessing 
would be elliptical-like, or highly irregular, rather than disk-like. 

The degree of the uncertainties involved in linking the spectroscopic and virial 
temperatures (e.g., Ciotti \& Pellegrini 2008), and some ambiguity in the definitions 
themselves (e.g., Voit 2005), should also be kept in mind. 
Moreover, the gas temperature estimated through the single thermal model is only a 
first approximation, and could be somewhat unreliable. The best fit to the full halo 
spectrum calls for two components with different temperatures and abundances (model E). 
It is then reasonable to associate the warmer ($kT_1 \sim 0.8$ keV) and metal-richer 
($Z_\alpha \sim 0.5$ solar) component with the chemically-evolved, starburst-injected 
gas, and to identify the cooler ($kT_2 \sim 0.25$ keV) and metal-poorer ($Z \equiv 0.1$ 
solar) component with the gravitationally-bound, pre-existing halo material. Also the 
energy dissipated during the merger process may contribute to the heating of the 
ambient gas. In this view, a fossil group nature for the system would not be strictly 
required, and the dark matter halo of NGC~6240 would be broadly consistent with 
just a pair of massive spiral galaxies. 

The lower temperature of $\sim$0.25 keV would be also in good agreement with 
the virial temperature $kT_\sigma \sim \mu m_p \sigma_*^2 \sim 0.3$ keV derived 
from the average stellar velocity dispersion, $\sigma_* \simeq 200$--220 km s$^{-1}$ 
(Engel et al. 2010b). This comparison is physically motivated only for early-type galaxies 
(e.g., Pellegrini 2011), which are actually believed to represent the ultimate evolutionary stage 
of a major merger such as NGC~6240. Although the potential of the nuclear dynamical 
mass is expected not to be dominant on the halo scales, the stellar radial surface brightness 
profile follows the $r^{1/4}$ law typical of elliptical galaxies out to $\sim$20--25 kpc 
(Bush et al. 2008). While some of the disk-like structures are still present, the whole remnant 
has apparently entered the final relaxation phase. 

\subsection{Metal Enrichment}
Another primary source of information to understand the nature of the X-ray halo is the 
metallicity pattern, including absolute and relative abundances and their dependence on 
galactocentric distance. The average $\alpha$-element to iron abundance ratio turns out 
to be generally supersolar all across the halo (Table~\ref{t7}). This is a key observable to trace 
the chemical evolution of the interstellar matter, due to the different production yields and 
typical time-scales of the various SNe populations involved. Synthesis models for type II SNe 
(i.e., core-collapsed massive stars) predict Si/Fe ratios up to $\sim$3--5 solar (e.g., Nomoto 
et al. 2006), while these drop to $\sim$0.5 solar for type Ia SNe (i.e., exploded white dwarfs 
in close binary systems). Evidence for a supersolar Si/Fe ratio, for instance, was found in 
galaxy mergers like the Antennae (Baldi et al. 2006a,b), and has been recently revealed in 
the central regions of young elliptical galaxies, which are the sites of the latest (a few tens 
of Myr), merger-induced star formation (Kim et al. 2012).

\begin{figure}
\includegraphics[width=15cm]{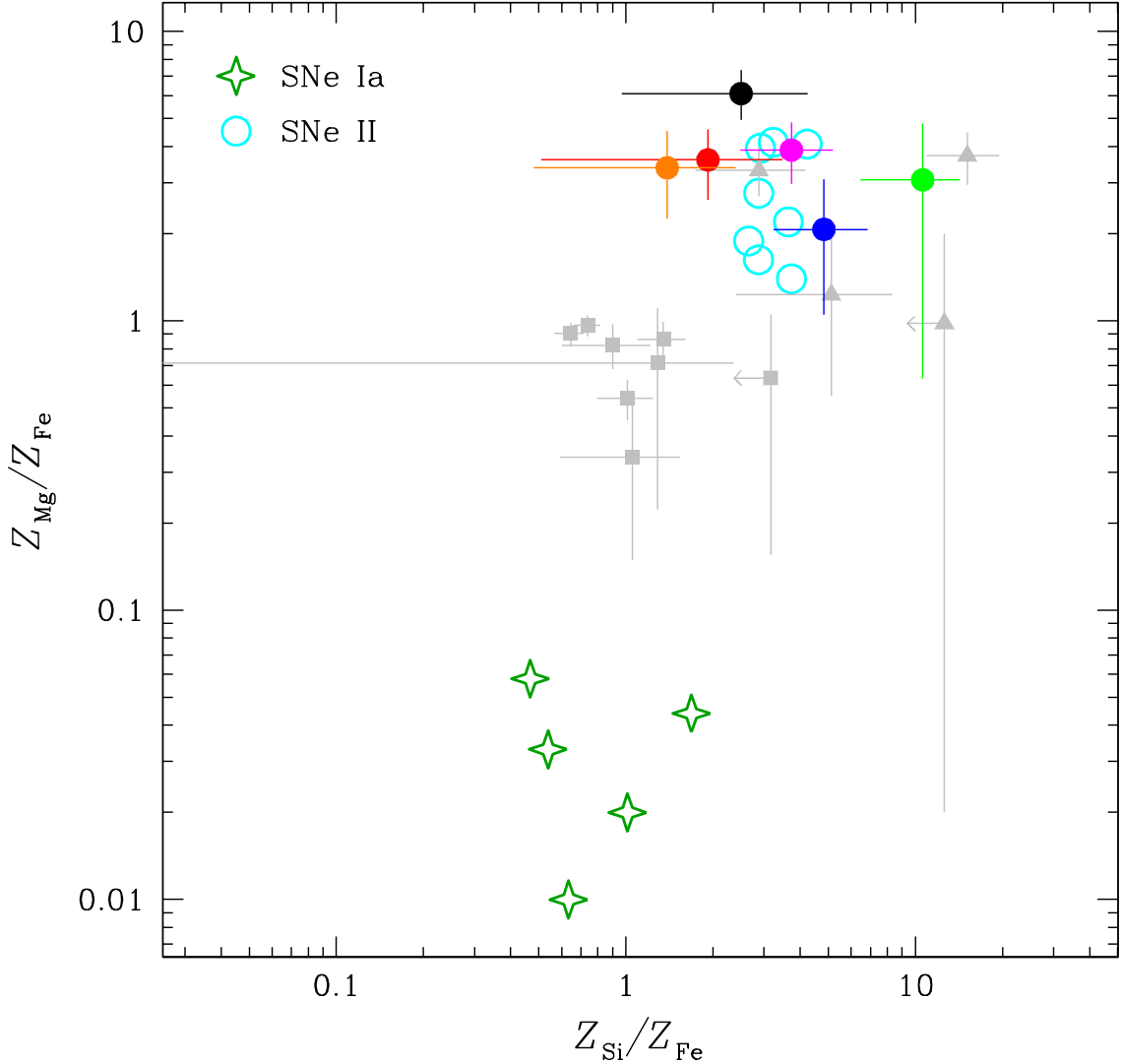}
\caption{Abundance ratio diagram, with $Z_\rmn{Mg}/Z_\rmn{Fe}$ plotted against 
$Z_\rmn{Si}/Z_\rmn{Fe}$ for the six halo subregions (dots, usual color code). The 
best-fit values are taken from model F, and the error bars are given at 1$\sigma$ 
to facilitate the comparison with measures from previous works (in grey): triangles 
represent the outflow, halo, central plane and outer plane regions of NGC~4490 
(Richings et al. 2010), while squares represent the three radial bins 
($r = 0$--30$\arcsec$, 30--60$\arcsec$, 60--120$\arcsec$) of the young ellipticals 
NGC~720 and NGC~3923, and the cores ($r = 0$--30$\arcsec$) of the quiescent 
early-type galaxies NGC~4472 and NGC~4649 (Kim et al. 2012; for all these sources 
the angular scale is $\approx$100--120~pc arcsec$^{-1}$, and the ratios have been 
converted to the solar standards adopted here). Reference values predicted by 
theoretical models are also shown: dark green stars for type Ia SNe, cyan circles for 
type II SNe (derived from Nomoto et al. 1997 and Nagataki \& Sato 1998, respectively).}
\label{en}
\end{figure}

In Fig.~\ref{en} the abundance ratios $Z_\rmn{Mg}/Z_\rmn{Fe}$ and $Z_\rmn{Si}/Z_\rmn{Fe}$ 
are plotted against each other as diagnostics of the enrichment history in the six azimuthal 
sectors of the halo. For comparison, the same values are shown for different regions of the 
disturbed spiral galaxy NGC~4490, which is interacting with the nearby irregular companion 
NGC~4485 (Richings et al. 2010), and of the elliptical galaxies analyzed by Kim et al. (2012). 
In such a diagram, the theoretical yields from various models of type Ia SNe (Nomoto et al. 
1997) and type II SNe (Nagataki \& Sato 1998) have a wide separation. All of the S1--S6 halo 
subregions definitely occupy the typical location of starburst environments, dominated by a 
young stellar population and type II SNe. From an evolutionary perspective, NGC~6240 should 
gradually move towards the intermediate position of early-type galaxies, following the takeover 
by type Ia SNe after the merger completion and the starburst fading. 

There is actually a possible caveat to the picture outlined above. By assuming in our fits a solar 
abundance for nickel, we obtain a nominal Ni/Fe ratio $\sim$7 times above solar in the full halo. 
Similar findings have been discussed over the past two/three decades for SN remnants 
(based on the strength of forbidden optical/near-IR [Ni~\textsc{ii}] lines; e.g., Bautista et al. 1996) 
and galaxy clusters (e.g., Dupke \& White 2000). On average, type Ia SNe are expected to have a 
much higher production yield for nickel with respect to type II SNe. A large relative abundance of 
nickel, if real, would imply a more complex enrichment history. As mentioned before, however, 
$Z_\rmn{Ni}$ can only be poorly constrained, due to its degeneracy with $Z_\rmn{Ne}$. Our 
assumption of a single thermal component might also introduce some bias. Indeed, by tying 
nickel and iron abundances in the two-temperature model D we achieve a statistically equivalent 
fit, with $\Delta \chi^2 < 0.5$, and just an increase of $Z_\rmn{Ne}$ to $\sim$0.6 solar. In 
general, the best-fitting abundance ratios from model D are slightly lower (Table~\ref{t7}), but 
still in excellent agreement with the predicted yields of type II SNe for all elements, nickel included.

Based on the above considerations, the whole halo of NGC~6240 appears to have experienced 
a significant metal enrichment by type II SNe. The constant abundance ratios in the inner and 
outer halo imply that the enrichment process has been uniform. As a consequence, the gas 
redistribution cannot be associated to recent starburst episodes, but has rather proceeded 
in the whole course of the interaction, whose timescale up to now is presumably $\sim$1 Gyr. 
Mergers of identical disk galaxies with massive stellar bulges are characterized by a single, 
intense peak of star formation during the nuclear coalescence (Springel et al. 2005; Cox et 
al. 2008). Entering this major burst, NGC~6240 is going to exceed the ULIRG luminosity threshold 
(Tacconi et al. 1999). The final stage is not imminent yet, since the two nuclei (which are the 
remnant of the 
progenitors' bulges) still have a projected separation of $\sim$0.8 kpc (1.5$\arcsec$; 
Max et al. 2007) and are possibly in the process of separating again (Tecza et al. 2000; 
Engel et al. 2010b). This notwithstanding, the identification of several tens of young circumnuclear 
star clusters, with ages in the range $\sim$5--15 Myr, confirms that enhanced star formation 
activity compared to isolated, quiescent galaxies is currently ongoing (Pasquali et al. 2003; 
Pollack et al. 2007). A steady star formation rate of this kind (61 $M_{\sun}$ yr$^{-1}$; Yun \& 
Carilli 2002) over the past $\sim$200 Myr (the dynamical age of the soft X-ray halo), perhaps 
involving the peripherical regions, can easily account for the energy and metallicity content 
of the hot gas component. The tentative decrease with radius of the absolute abundances 
hints at the gentle mixing of the starburst-driven outflows with a pre-existing halo medium, 
as also suggested by the best-fitting spectral model E. If no dilution effect were present and 
the dispersion of metal-rich gas were highly efficient, the observed metallicity gradient would 
be positive, with supersolar abundances at the larger distances (Cox et al. 2006). 

The fate of the starburst-processed gas is not clear. Some merger 
remnants, and young ellipticals bearing the signatures of recent 
interactions, are under-luminous in the X-rays with respect to 
relaxed early-type galaxies (Fabbiano \& Schweizer 1995; Read 
\& Ponman 1998), while different mechanisms of halo 
regeneration have been proposed (O'Sullivan et al. 2001b). If the 
wind temperature is much larger than the virial temperature, most 
of the hot gas is not bound and will eventually escape the system, 
becoming undetectable in the X-rays due to the density drop. In 
NGC~6240, the single-temperature cooling time of a few Gyr implies 
that the halo luminosity will not be fully extinguished after the merger 
conclusion. In the more complex scenarios, the different gas phases 
might have much shorter cooling times, depending on their degree of 
mixing. According to model E, unless the filling factors are very small 
($\eta \ll 0.1$), the radiative cooling of the warmer gas component will 
take several hundreds of Myr at least. Provided that some favorable 
conditions are met (e.g., the ultimate AGN feedback is not dramatically 
powerful, the cooler gas component exerts a sufficient drag, and the 
gravitational potential well is deep enough), NGC~6240 could be a 
case of major merger with no complete blow-out. 

\begin{table}
\caption{Relative Abundances of $\alpha$-elements with respect to Iron.} 
\label{t7}
\begin{tabular}{cccccc}
\hline \hline
Reg & Mod & $Z_\rmn{O}/Z_\rmn{Fe}$ & $Z_\rmn{Ne}/Z_\rmn{Fe}$ & $Z_\rmn{Mg}/Z_\rmn{Fe}$ & $Z_\rmn{Si}/Z_\rmn{Fe}$ \\ 
\hline 
FH & A & 4.44$^{+1.75}_{-1.51}$ & 5.07$^{+1.42}_{-1.41}$ & 3.83$^{+0.91}_{-0.94}$ & 4.13$^{+1.70}_{-1.52}$ \\
 & B & 4.58$^{+1.73}_{-1.43}$ & 5.25$^{+1.41}_{-1.33}$ & 3.90$^{+0.90}_{-0.92}$ & 3.95$^{+1.55}_{-1.43}$ \\
 & C & 4.63$^{+1.38}_{-1.18}$ & 5.10$^{+1.22}_{-1.19}$ & 3.80$^{+0.77}_{-0.77}$ & 3.85$^{+1.17}_{-1.04}$ \\
 & D & 2.44$^{+1.81}_{-1.00}$ & 2.13$^{+3.08}_{-1.58}$ & 3.73$^{+1.09}_{-1.01}$ & 3.28$^{+1.93}_{-1.25}$\smallskip \\
S1 & F & 5.34$^{+3.72}_{-2.65}$ & 5.77$^{+3.08}_{-2.77}$ & 6.10$^{+2.13}_{-1.88}$ & 2.50$^{+3.04}_{-2.47}$ \\
 & G & 5.29$^{+2.90}_{-2.39}$ & 4.52$^{+2.69}_{-2.40}$ & 5.80$^{+2.09}_{-1.92}$ & 2.36$^{+2.70}_{-2.36}$\smallskip \\ 
S2 & F & 2.19$^{+2.64}_{-1.91}$ & 3.54$^{+2.75}_{-3.20}$ & 2.07$^{+1.70}_{-1.71}$ & 4.83$^{+3.65}_{-2.47}$ \\
 & G & 3.68$^{+2.31}_{-1.95}$ & 4.46$^{+2.53}_{-2.31}$ & 2.06$^{+1.59}_{-1.45}$ & 4.53$^{+2.54}_{-2.24}$\smallskip \\
S3 & F & 4.80$^{+3.07}_{-2.29}$ & 3.77$^{+2.47}_{-2.40}$ & 3.89$^{+1.61}_{-1.51}$ & 3.73$^{+2.55}_{-1.98}$ \\
 & G & 5.60$^{+2.64}_{-2.16}$ & 3.20$^{+2.29}_{-2.11}$ & 3.77$^{+1.71}_{-1.57}$ & 3.91$^{+2.38}_{-2.09}$\smallskip \\
S4 & F & 3.87$^{+3.38}_{-2.22}$ & 4.05$^{+2.49}_{-2.42}$ & 3.60$^{+1.66}_{-1.62}$ & 1.92$^{+2.69}_{-1.92}$ \\
 & G & 5.33$^{+2.96}_{-2.44}$ & 3.59$^{+2.57}_{-2.32}$ & 3.20$^{+1.88}_{-1.74}$ & 1.60$^{+2.69}_{-1.60}$\smallskip \\
S5 & F & 5.60$^{+3.35}_{-2.77}$ & 1.67$^{+4.25}_{-1.67}$ & 3.38$^{+1.95}_{-1.85}$ & 1.39$^{+1.74}_{-1.39}$ \\
 & G & 3.01$^{+2.14}_{-1.80}$ & 7.67$^{+2.96}_{-2.61}$ & 4.96$^{+1.82}_{-1.66}$ & 2.91$^{+2.29}_{-2.02}$\smallskip \\
S6 & F & 7.89$^{+7.47}_{-4.22}$ & 9.59$^{+5.72}_{-4.21}$ & 3.07$^{+2.91}_{-3.07}$ & 10.6$^{+6.5}_{-5.6}$ \\
 & G & 5.67$^{+2.88}_{-2.36}$ & 7.44$^{+3.04}_{-2.63}$ & 3.20$^{+1.82}_{-1.68}$ & 7.67$^{+3.12}_{-2.64}$ \\     
\hline
\end{tabular}
\flushleft
\small{\textit{Note.} Since the errors on the individual elements are asymmetric and strongly 
correlated, these cannot be added in quadrature to obtain the confidence ranges for ratios, 
which have been computed, instead, by introducing a dummy Gaussian component with zero 
normalization, and linking a multiplicative factor to the abundance ratio.}
\end{table}

\section{Summary and Conclusions}
We have presented an X-ray spectral and imaging analysis of the ultraluminous infrared 
galaxy merger NGC~6240, based on the combined $\sim$182-ks exposure from a pair 
of new (145 ks) and archival (37 ks) \textit{Chandra} observations. NGC~6240 is a very 
complex system at all the physical scales, sporting a heavily-obscured, dual AGN in 
the central kpc, a circumnuclear environment swept by powerful starburst-driven winds, 
and a spectacular halo of hot gas, which definitely make of this source one of the most 
intriguing extragalactic targets. In this work, we have focused on the diffuse, soft X-ray 
emission at $r > 15\arcsec$ ($\sim$7.5 kpc) from the visual centroid, coincident with 
the southern nucleus. Our main results can be summarized as follows:

\begin{enumerate}
\item In the soft X-rays, the surface brightness of NGC~6240 exceeds the background at the 
3$\sigma$ confidence level out to an average radius of $\sim$50 kpc. This huge halo shows a 
slightly elongated, diamond-like shape, with nearly perpendicular directions of maximum and 
minimum extension and a projected size at full length of $\sim$110$\times$80 kpc. The global 
morphology is clumpy and filamentary throughout this range. 

\item The X-ray emission from the full halo is consistent with a single-temperature gas 
heated at $\sim$0.65 keV (7.5 million K), with a typical density of a few $\times$10$^{-3}$ 
cm$^{-3}$ and a total mass of 1.3$\times$10$^{10} M_{\sun}$. Given the estimated thermal 
energy content of $\sim$5$\times$10$^{58}$ erg, the intrinsic 0.4--2.5 keV luminosity 
of 4$\times$10$^{41}$ erg s$^{-1}$ implies a cooling time larger than a Gyr, under the 
assumption of a filling factor $\eta \sim 1$. 

\item The spatially-resolved spectral analysis, feasible thanks to the adequate number of 
net counts available, reveals limited variations, with no apparent correlation of the gas 
physical parameters with the main morphological features. In the radial direction, the 
inferred properties clearly differ from those expected for an adiabatic outflow, being the 
temperature virtually constant and the density slowly decreasing. 

\item Absolute abundances are generally subsolar, with possible gradients among the 
azimuthal sectors, whose statistical significance is anyway marginal. In spite of the large 
error bars, a tentative metallicity drop with radius is found with a cumulative 2$\sigma$ 
confidence. Abundance ratios between the main $\alpha$-elements and iron are several 
times the solar value, calling for a metal enrichment dominated by type II SNe. 
\end{enumerate}

Based on the luminosity of its X-ray halo, NGC~6240 might even be the remnant of a small 
group of galaxies, on its way to becoming a massive elliptical. The gas heating can be ascribed 
to different mechanisms, including the dissipation of kinetic energy during the galactic collision, 
the fast shocks within a starburst-driven superwind, the virialization by a group-like 
gravitational potential. When examined individually, none of the latter interpretations accounts 
in a self-consistent manner for the observational evidence in its entirety. Although the application 
of a simple thermal model provides useful information on the basic properties of the halo, the 
actual physical conditions are likely much more complex. Indeed, the best-fitting spectral model 
consists of a multi-phase gas with different temperature and metallicity, and suggests the mixing 
of two primary gas components. In our favorite scenario, the warmer one ($kT \sim 0.8$ keV) 
has been chemically-contaminated ($Z_\alpha \sim 0.5$ solar) by a regular, widespread star 
formation activity over the past $\sim$200 Myr, following the dynamical interaction of the parent 
galaxies; the cooler one ($kT \sim 0.25$ keV) is related to the metal-poor ($Z \sim 0.1$ solar), 
pre-existing ambient material, which has been heated up to X-ray emitting temperatures due to 
either the merger-induced dissipation and shocks or the binding energy of the dark matter halo. 
The present data quality does not allow us to probe this picture in more depth. Subject to such a 
high spatial resolution, a further improvement (by at least a factor of $\sim$5) in the amount of 
net counts available would be needed to clearly disentangle the gas components of different 
temperature, and to significantly constrain any abundance gradient within the halo. While the 
system is already relaxing, and the final coalescence of the two nuclei will ultimately lead to the 
formation of a young elliptical galaxy, it is still unclear how much of the hot, diffuse gas can be 
retained in the form of an X-ray bright, hydrostatic halo. 

In conclusion, this work confirms NGC~6240 as one of the most striking sources in the local 
Universe, whose study delivers plenty of challenging information for both theoretical models 
and numerical simulations, with far-reaching implications in all the fields of galaxy formation 
and evolution. 


\textit{Acknowledgments}. This research is supported by NASA grant GO1-12123X 
(PI: G. Fabbiano). We thank the anonymous referee for a careful reading, and constructive 
comments that helped improving the clarity of this paper. We acknowledge support 
from the \textit{Chandra X-ray Center} (CXC), which is operated by the Smithsonian 
Astrophysical Observatory (SAO) for and on behalf of NASA under contract NAS8-03060. 
We have made use of software provided by the CXC in the application packages \textsc{ciao} 
and \textit{Sherpa}, and of SAOImage DS9, developed by SAO. The X-ray data were retrieved 
from the \textit{Chandra} Data Archive, and the \textit{HST} image was obtained from the MAST 
data archive at the Space Telescope Science Institute. STScI is operated by the Association of 
Universities for Research in Astronomy, Inc., under NASA contract NAS5-26555. 

\section*{Appendix}

We summarize below the spectral models employed in this work, and their  
underlying assumptions in the application to the different halo regions. The 
basic form involves thermal emission from hot gas in collisional ionization 
equilibrium, modified by Galactic absorption only. We also probed the possible 
presence of neutral column density local to the source, and the contribution 
of an additional power-law or thermal component. Abundances of Fe, O, Ne, 
Mg, and Si are allowed to be different fractions of solar values. \\
A -- \texttt{wabs*zwabs*(vapec+powerlaw)} -- Full halo spectrum, grouped to a 
4$\sigma$ significance per energy channel and fitted with the $\chi^2$ statistic. \\
B -- \texttt{wabs*zwabs*(vapec+powerlaw)} -- \textit{Ungrouped} (with at least 
one count per bin) spectra of the full, inner and outer halo regions, fitted with 
the $C$-statistic. \\
C -- \texttt{wabs*vapec} -- Ungrouped spectra of the azimuthal regions S1--S6, 
fitted simultaneously with both temperatures and abundances tied, and of the 
radial regions H1--H5, fitted simultaneously in the spatially-resolved analysis 
with temperatures free and abundances tied. \\
D -- \texttt{wabs*(vapec+vapec)} -- Grouped spectrum of the full halo, with 
abundances tied between the two thermal components. \\
E -- \texttt{wabs*(vapec+vapec)} -- Grouped spectrum of the full halo, with abundance 
set to 0.1 solar for all the elements in the low-temperature component. \\
F -- \texttt{wabs*vapec} -- Ungrouped spectra of the individual S1--S6 regions,  
fitted independently in the spatially-resolved azimuthal analysis. \\
G -- \texttt{wabs*vapec} --  Ungrouped S1--S6 spectra, fitted simultaneously with 
a single parameter of interest free to vary, in turn, over the different regions, and 
tied otherwise. 



\end{document}